\documentclass[final,leqno]{siamltex}
\usepackage{array}
\usepackage{amsmath}
\usepackage{amssymb}
\usepackage{array}
\usepackage{float}
\usepackage{color}
\usepackage{graphicx}
\usepackage{epsfig}
\usepackage{graphics}
\usepackage{pstricks}

\makeatletter\@addtoreset {equation}{section}\makeatother

\author{Jamie M. Foster$^{1}$, Peter Gysbers$^{2}$, John R. King$^3$ \& Dmitry E. Pelinovsky$^2$ \\
    {\em $^1$}\\
    {\em $^2$}\\
    {\em $^3$}
    }

\title{Bifurcations of self-similar solutions for reversing interfaces in the slow diffusion equation with strong absorption}
\author{
J. M. Foster\thanks{Department of Mathematics, University of Portsmouth, Portsmouth, UK, PO1 2UP (jamie.foster@port.ac.uk).}
\and P. Gysbers\thanks{Department of Physics \& Astronomy, McMaster University, Hamilton ON, Canada, L8S 4K1.}
\and J. R. King\thanks{School of Mathematical Sciences, Nottingham University, Nottingham, UK, NG7 2RD.}
\and D. E Pelinovsky\thanks{Department of Mathematics \& Statistics, McMaster University, Hamilton ON, Canada, L8S 4K1.}
}

\date{\today}
\begin{document}
\maketitle
\begin{abstract}
Bifurcations of self-similar solutions for reversing interfaces are studied in the slow diffusion equation with strong absorption.
The self-similar solutions bifurcate from the time-independent solutions for standing interfaces.
We show that such bifurcations occur at the bifurcation points, at which
the confluent hypergeometric functions satisfying Kummer's differential equation
is truncated into a finite polynomial. A two-scale asymptotic method is employed to obtain the asymptotic dependencies
of the self-similar reversing interfaces near the bifurcation points. The asymptotic results are shown to be in
excellent agreement with numerical computations.
\end{abstract}


\thispagestyle{plain}

\section{Introduction}

We address reversing interfaces in the following slow diffusion equation with strong absorption
\begin{equation}
\label{heat}
\frac{\partial h}{\partial t} = \frac{\partial}{\partial x} \left( h^m \frac{\partial h}{\partial x} \right) - h^n,
\end{equation}
where $h(x,t)$ is a positive function on a compact support, \emph{e.g.}, a concentration of some species, and
$x$ and $t$ denote space and time, respectively. Restricting the exponents to the ranges $m > 0$ and $n < 1$
limits our interest to the case of slow diffusion \cite{Herrero} and strong absorption \cite{Kalash,Kersner2,Kersner},
respectively. The restriction $m>0$ implies that the edges of the compact support
propagate with a finite speed \cite{Galaktionov}, whilst for $n<1$ compactly supported solutions extinct in a finite time \cite{Chen}.
These two results suggest that, for certain choices of initial data which lead to an initial expansion
of the compact support, reversing of interfaces can occur. Here, we use the term ``reversing of interfaces''
to describe a scenario in which an advancing interface gives way to a receding interface; the term ``anti-reversing'' of an interface describes the converse, \emph{i.e.}, a receding interfaces giving way to an advancing one. Such scenarios have been examined in this range of exponents previously in \cite{Foster,FosterPel}. For the special case when $m+n=2$ it has been shown, in \cite{Galaktionov2}, that solutions can exhibit reversing interfaces but cannot display a ``waiting time'' where an interface remains static for some finite time. The behaviour of solutions local to the extinction time has also been examined in the limiting case when $m+n=1$ in \cite{Gala3,Gala4} and in the case $m>1$ and $n<1$ in \cite{pablo}.

The slow diffusion equation with strong absorption, (\ref{heat}), is relevant in a wide variety of different physical processes and can be used as a model for: (i) the slow spreading of a slender viscous film over a horizontal plate subject to the action of gravity and a constant
evaporation rate \cite{Acton} (when $m = 3$ and $n = 0$); (ii) the dispersion of a biological population subject to a constant
death-rate \cite{popn} (when $m = 2$ and $n = 0$); (iii) nonlinear heat conduction along a rod with a constant
rate of heat loss \cite{Herrero} (when $m = 4$ and $n = 0$), and; (iv) fluid flows in porous media with a drainage rate driven by gravity or background flows \cite{Aronson,PWH} (when $m = 1$ and
either $n = 1$ or $n = 0$).

After selecting the origin of the spatial and temporal coordinates such that the region of positive $h$ lies in $x>0$ and the reversing or anti-reversing event occurs at $t=0$ (at which time the interface is located at $x=0$),
a plausible local behaviour of interfaces is provided by the self-similar solutions
in the form suggested in \cite{Foster},
\begin{equation}
\label{reduction}
h(x,t) = \left( \pm t \right)^{\frac{1}{1-n}} \; H_{\pm}(\xi), \quad \xi = x (\pm t)^{-\frac{m+1-n}{2(1-n)}}, \quad \pm t > 0,
\end{equation}
where the functions $H_{\pm}$ satisfy the following second-order ordinary differential equations (ODEs):
\begin{equation}
\label{ode}
\frac{d}{d \xi} \left( H_-^m \frac{d H_-}{d \xi} \right) - \frac{m+1-n}{2 (1-n)} \, \xi \frac{d H_-}{d \xi} = H_-^n - \frac{1}{1-n} H_-
\end{equation}
and
\begin{equation}
\label{ode-plus}
\frac{d}{d \xi} \left( H_+^m \frac{d H_+}{d \xi} \right) + \frac{m+1-n}{2 (1-n)} \, \xi \frac{d H_+}{d \xi} = H_+^n + \frac{1}{1-n} H_+.
\end{equation}
The mass preserving free-boundary conditions for (\ref{heat}) are
\begin{equation} \label{noleaks}
h \to 0^+, \quad \frac{ds_{\pm}}{dt} = h^m \frac{\partial h}{\partial x} + h^n \left( \frac{\partial h}{\partial x} \right)^{-1} \quad \mbox{as} \quad x \to s_{\pm}(t)^+,
\end{equation}
where $x=s_{\pm}(t)$ is the location of the free boundary for positive and negative time respectively.
The form of the self-similar solution (\ref{reduction}) implies that the interface after and before
a reversing or anti-reversing event is located at the positions given by
\begin{equation} \label{frontbeh}
s_{\pm}(t) = \hat{\xi}_{\pm} (\pm t)^{\frac{m+1-n}{2(1-n)}},
\end{equation}
where $\hat{\xi}_{\pm}$ are both constants with $\hat{\xi}_+$ being relevant for $t>0$
whilst $\hat{\xi}_-$ is relevant for $t<0$. Owing to (\ref{frontbeh}), in addition to requiring
$m > 0$ and $n < 1$, we are also restricted to $m + n > 1$ so that $s_{\pm}(t)$ has
a physically reasonable behaviour in time with $\lim_{t \to \pm 0} \dot{s}_{\pm}(t) = 0$.
For technical reasons described below (\ref{indices}), we also restrict the range
of $n$ to $-1 \leq n < 1$.

The conditions (\ref{noleaks}) and (\ref{frontbeh})
imply that solutions to (\ref{ode}) and (\ref{ode-plus}) are required to satisfy
\begin{equation} \label{pressure}
H_\pm \to 0^+, \quad H^m_\pm \frac{dH_{\pm}}{d\xi} \to 0^+ \quad \mbox{as} \quad \xi \to \hat{\xi}_{\pm}^+.
\end{equation}
For reasons that will become clear shortly the far-field condition
\begin{equation} \label{ocean} 
H_\pm \sim \left(\frac{\xi}{A} \right)^{\frac{2}{m+1-n}} \quad \mbox{as} \quad \xi \to +\infty
\end{equation}
completes the specification of the relevant boundary value problems for the system (\ref{ode})-(\ref{ode-plus}).
The constant $A>0$ is determined from solving equation (\ref{ode}) and the same $A$ is prescribed
while solving equation (\ref{ode-plus}). Requiring identical far-field behaviours in the solution
of both (\ref{ode}) and (\ref{ode-plus}) is tantamount to ensuring continuity of $h$ across $t=0$
with
\begin{equation}
\lim_{t \to 0} h(x,t) = \left( \frac{x}{A} \right)^{\frac{2}{m+1-n}}.
\end{equation}

Existence of suitable solutions to the boundary-value problem (\ref{ode}), (\ref{ode-plus}),
(\ref{pressure}), and (\ref{ocean}) was first suggested in \cite{Foster}. Later, this was elaborated
in \cite{FosterPel} with an analytic shooting method that made use of invariant manifold theory for
dynamical systems in appropriately rescaled variables near the small and large values of $H_{\pm}$.

We note the existence of an exact solution to (\ref{ode})--(\ref{ode-plus}) in the form
\begin{equation}
\label{exact-solution}
H_{\pm}(\xi) = \left( \frac{(m+1-n)^2}{2(m+n+1)} \xi^2 \right)^{\frac{1}{m+1-n}}.
\end{equation}
The conditions (\ref{pressure}) and (\ref{ocean}) are satisfied with $\hat{\xi}_{\pm} = 0$
and $A = A_Q$, where
\begin{equation} \label{aq_def}
A_Q := \left(\frac{2(m+1+n)}{(m+1-n)^2} \right)^{1/2}.
\end{equation}
It is straightforward to verify that in the original spatial and temporal variables given by (\ref{reduction})
the exact solution (\ref{exact-solution}) corresponds to a time-independent solution
to the slow diffusion equation (\ref{heat}) given by
\begin{equation}
\label{exact-solution-heat}
h(x) = \left( \frac{(m+1-n)^2}{2(m+1+n)} x^2 \right)^{\frac{1}{m+1-n}}.
\end{equation}
Hence, the interface is static for the exact solution in (\ref{exact-solution}) or (\ref{exact-solution-heat}).
Although this solution does not constitute a reversing or anti-reversing interface solution, it does play a central role
in the bifurcation analysis. Henceforth we refer to (\ref{exact-solution}) as the {\em primary branch}
of self-similar solutions to (\ref{ode}) and (\ref{ode-plus}).

Given that there is a key difference between the boundary--value problems for $H_-$ and $H_+$,
we shall provide local analysis of the asymptotic expansions as $\xi \to \hat{\xi}_{\pm}^+$.
Two types of the leading-order balance may occur here. The first possibility is
the usual balance for porous-medium equations, in which the absorption term, $-h^n$, is negligible, \emph{i.e.}
\begin{equation} \label{sorder}
\frac{d}{d\xi} \left( H_\pm^m \frac{dH_{\pm}}{d\xi} \right) \sim \mp \frac{m+1-n}{2(1-n)} \hat{\xi}_{\pm} \frac{dH_{\pm}}{d\xi}.
\end{equation}
In view of the boundary conditions (\ref{pressure}), the balance (\ref{sorder})
is valid for $\pm \hat{\xi}_{\pm} < 0$ and yields the following local behaviour
\begin{equation} \label{onebeh}
H_{\pm} \sim \left( \mp \frac{m (m+1-n) \hat{\xi}_{\pm}}{2 (1-n)} (\xi - \hat{\xi}_{\pm}) \right)^{1/m}
\quad \mbox{\rm as} \quad \xi \to \hat{\xi}_{\pm}.
\end{equation}
The second possibility arises when the diffusion term is negligible, \emph{i.e.}
\begin{equation} \label{forder}
\pm \frac{m+1-n}{2(1-n)} \hat{\xi}_{\pm} \frac{dH_{\pm}}{d\xi} \sim H_\pm^n.
\end{equation}
The balance (\ref{forder}) is valid for $\pm \hat{\xi}_{\pm} > 0$ and yields the following local behaviour
\begin{equation} \label{genes}
H_{\pm}(\xi) \sim \left[ \pm \frac{2 (1-n)^2}{(m+1-n) \hat{\xi}_{\pm}} (\xi - \hat{\xi}_{\pm}) \right]^{1/(1-n)}
\quad \mbox{\rm as} \quad \xi \to \hat{\xi}_{\pm}.
\end{equation}
The asymptotic behaviours (\ref{onebeh}) and (\ref{genes}) were
proven rigorously in \cite{FosterPel} by using rescaling and dynamical system methods.

Unlike (\ref{sorder}), equation (\ref{forder}) is of first order and the second degree of freedom should be checked in the usual way by the Liouville-Green (JWKB) method, whereby linearisation about (\ref{forder}) implies a contribution
\begin{equation}
\exp \left( \frac{|\hat{\xi}_\pm|}{2} \left( \frac{(m+1-n)|\hat{\xi}_\pm|}{2(1-n)^2} \right)^{\frac{m}{1-n}} \left( \xi - \hat{\xi}_\pm \right)^{-\frac{m+n-1}{1-n}} \right)
\end{equation}
to the local expansion and is therefore inadmissible, \emph{i.e.}, each
of the balances (\ref{sorder}) and (\ref{forder}) contain only one degree of freedom, namely $\hat{\xi}_{\pm}$.

As $\xi \to +\infty$ the behaviour (\ref{ocean}) arises from the balance
\begin{equation} \label{sea}
\pm \frac{m+1-n}{2(1-n)}\xi \frac{dH_\pm}{d\xi} \sim \pm \frac{1}{1-n} H_\pm.
\end{equation}
Equation (\ref{sea}) is again of first order and linearising about (\ref{ocean})
to reinstate the second possible degree of freedom leads in this case to
\begin{equation}
\exp \left( \mp \left( \frac{m+1-n}{2(1-n)} A^{\frac{m}{m+1-n}} \right)^2  \xi^{\frac{2(1-n)}{m+1-n}} \right)
\end{equation}
for $H_\pm$ respectively. Therefore, the second solution is inadmissible for $H_-$ (lower sign),
leaving a single degree of freedom (namely $A$) as $\xi \to +\infty$, while
it is admissible for $H_+$ (upper sign). Again,
the behaviour (\ref{ocean}) was justified in \cite{FosterPel} by using rescaling and dynamical system methods.

In summary, ODEs (\ref{ode}) and (\ref{ode-plus}) are to be solved subject to
the boundary conditions (\ref{pressure}) and hence (\ref{onebeh}) or (\ref{genes}) as $\xi \to \hat{\xi}_{\pm}^+$,
where $\hat{\xi}_{\pm}$ is determined as a part of the solution $H_{\pm}$. In terms of the dynamics of
the PDE (\ref{heat}), the local behaviour of the solutions switches from
\begin{equation}
\label{first-wave}
h(x,t) \sim \left( -m\dot{s}(x-s)\right)^{\frac{1}{m}} \quad \mbox{with} \quad \dot{s}<0,
\end{equation}
as in (\ref{onebeh}), to
\begin{equation}
\label{second-wave}
h(x,t) \sim \left( \frac{1-n}{\dot{s}} (x-s) \right)^{\frac{1}{1-n}} \quad \mbox{with} \quad \dot{s}>0,
\end{equation}
as in (\ref{genes}) (or vice versa). The parameter $A$ in the boundary condition
(\ref{ocean}) is determined as a part of the solution $H_-$ by using a single-parameter shooting
from either $\xi \to \hat{\xi}_-^+$ or $\xi \to +\infty$. The value
of $A$ is prescribed as a part of the solution $H_+$.

\begin{figure}          \centering
\includegraphics[width=0.48\textwidth]{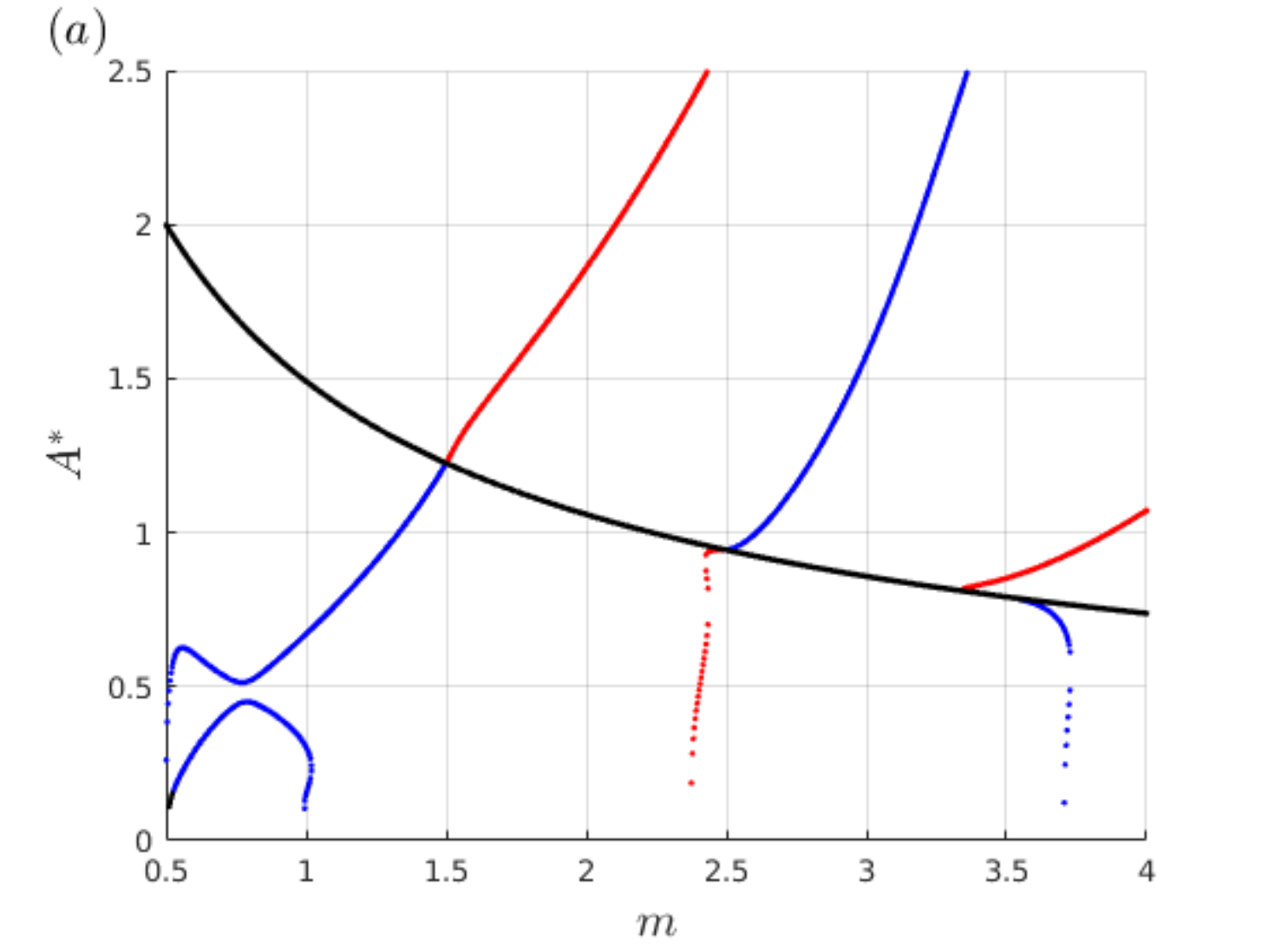}
\includegraphics[width=0.48\textwidth]{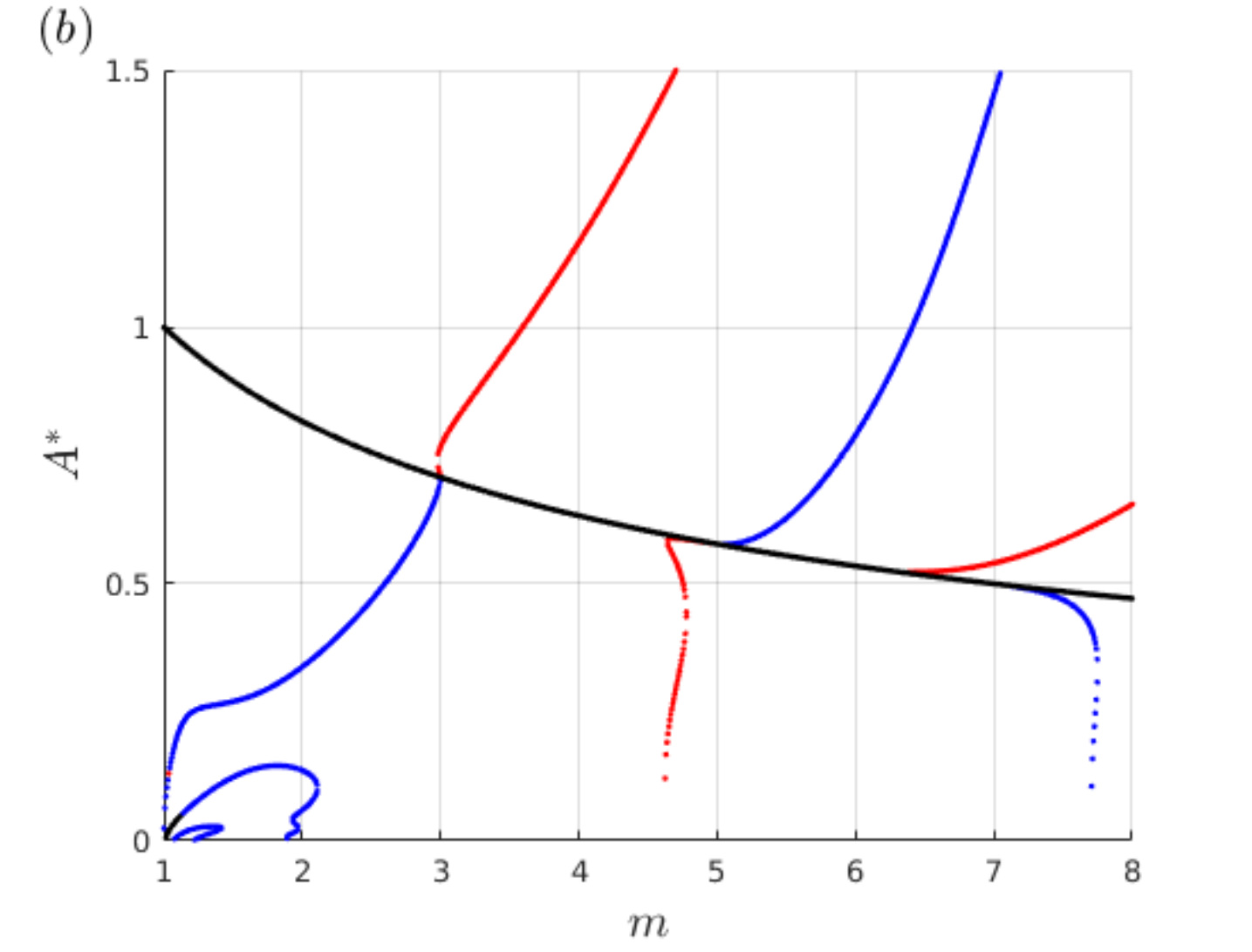}
\includegraphics[width=0.48\textwidth]{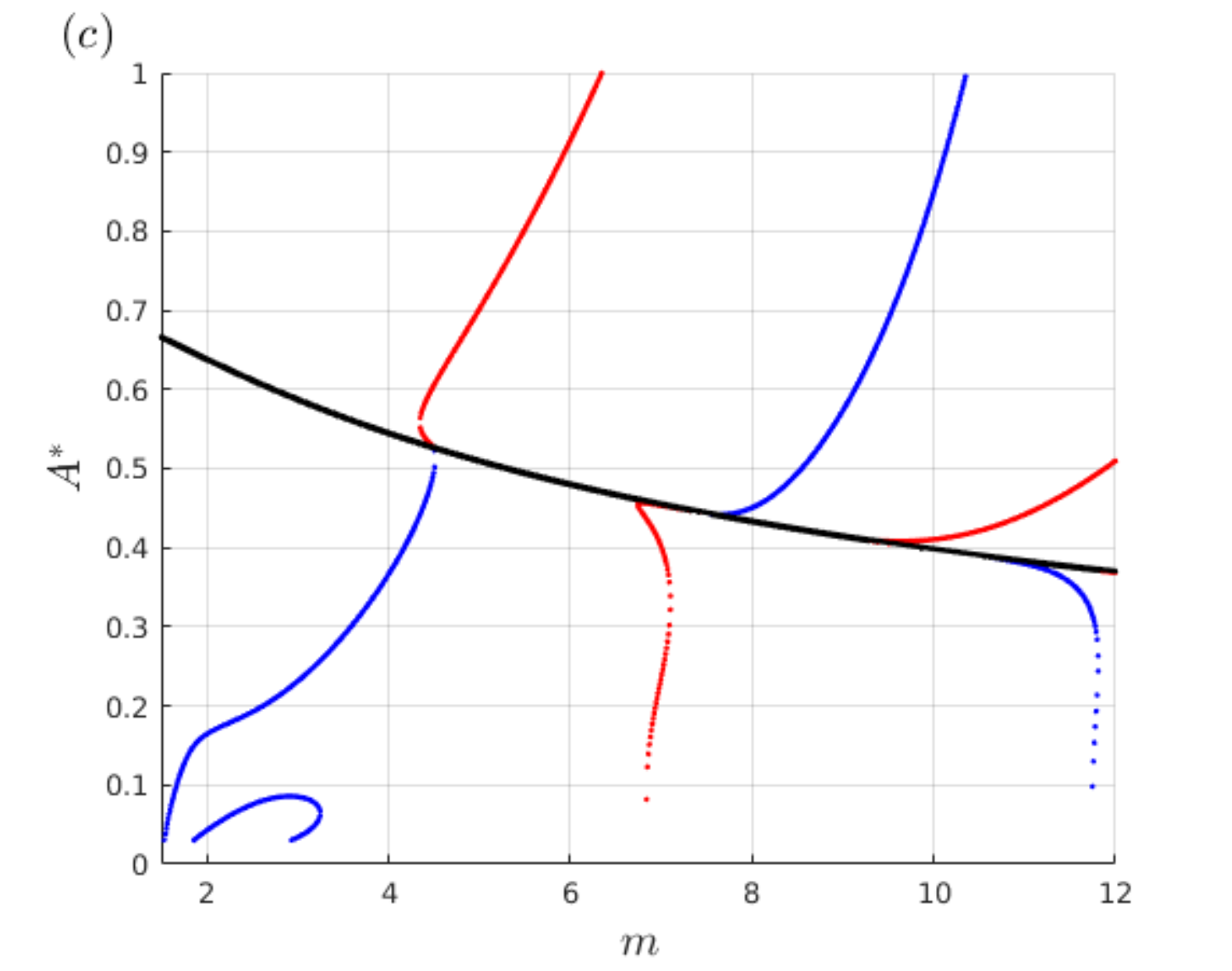}
\includegraphics[width=0.48\textwidth]{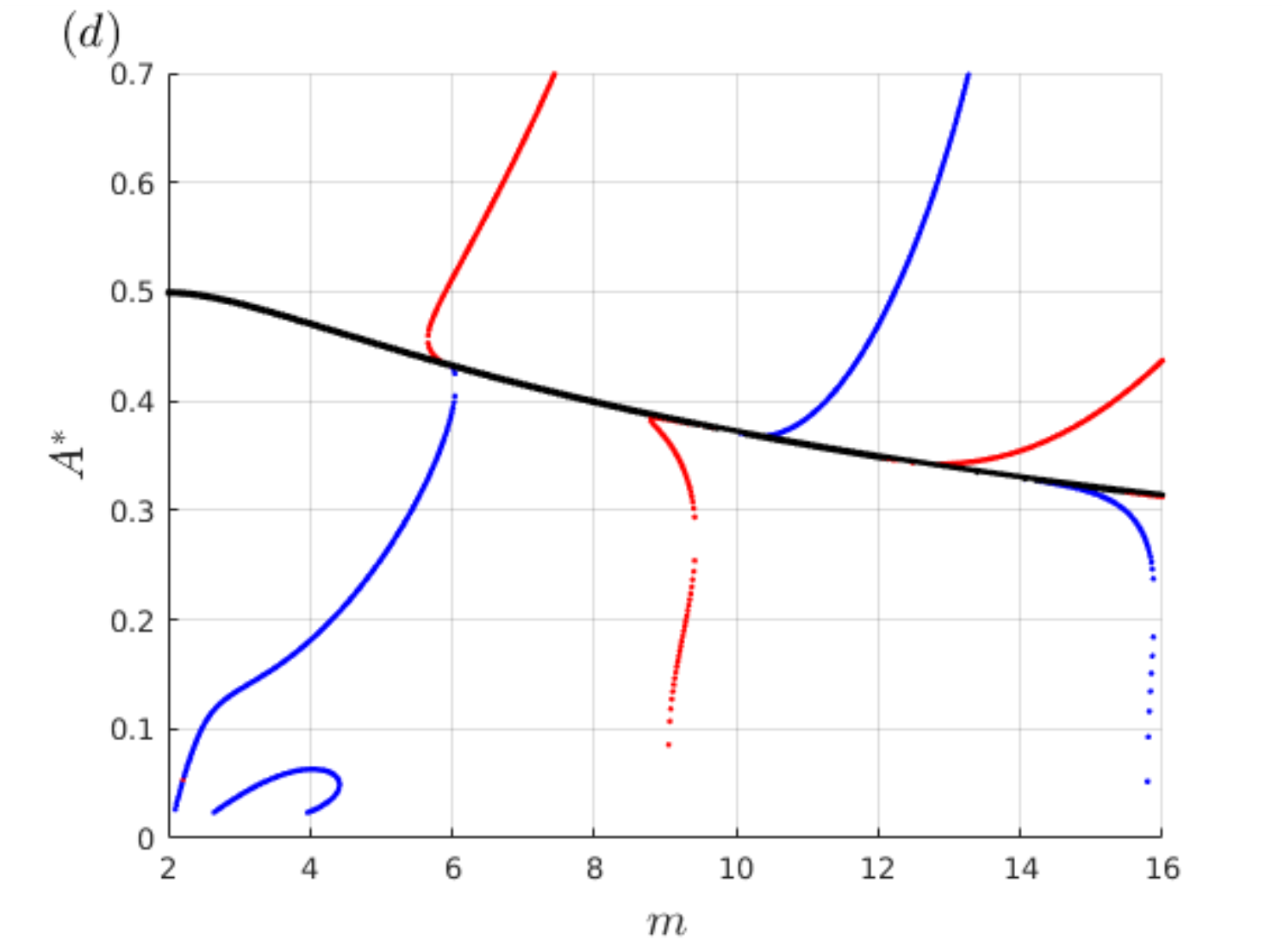}
\caption{The special values of $A=A^*$ (as a function of $m$) in the behaviour (\ref{ocean})
that correspond to a solution of the ODE (\ref{ode}) with the near field behaviours (\ref{onebeh}) and (\ref{genes}). The red, blue and black curves indicates values of $A^*$ that are associated
with solutions with $\hat{\xi}_- > 0$, $\hat{\xi}_- < 0$, and $\hat{\xi}_- = 0$ respectively. Panels (a)-(d) show the results for values of $n=1/2,0,-1/2,-1$ respectively.}
\label{all-sols-negative-t}
\end{figure}

A numerical shooting method was developed in \cite{FosterPel} for the case $n = 0$ to connect the near-field
and far-field behaviours for (\ref{ode}). The connection is possible only for some isolated values
of $A$ denoted by $A^*$. This shooting method was used here to obtain a diagram
of some possible self-similar solutions in the $(A,m)$-plane for few selected
values of $n$, see figure \ref{all-sols-negative-t}.
For each point on these diagrams, there is a unique value of $\hat{\xi}_-$
which is positive on the red curves, negative on blue curves, and zero on black curves.
A short summary of how the plots in figure \ref{all-sols-negative-t} were produced is
given in \S\ref{numerics} and full details can be found in \cite{FosterPel}.

In addition to providing a scheme for connection of the relevant behaviours for the ODE (\ref{ode})
the work in \cite{FosterPel} also demonstrated that all values of $A$ in the behaviour (\ref{ocean})
lead to a valid solution to the ODE (\ref{ode-plus}) for $H_+$. Moreover, it was shown that $\hat{\xi}_+$ is
a monotonic increasing function of $A$ with the following properties: if $A<A_Q$ then $\hat{\xi}_+<0$,
if $A>A_Q$ then $\hat{\xi}_+>0$ whereas if $A=A_Q$ then $\hat{\xi}_+=0$. Thus, if a valid solution
to (\ref{ode}) can be found, a related solution to (\ref{ode-plus}) can always be constructed for
the same value of $A=A^*$.

The rather exotic patterns visible on figure \ref{all-sols-negative-t} depict the existence
of bifurcating solutions from the black curve that corresponds to the case $\hat{\xi}_- = 0$,
and hence to the exact solution (\ref{exact-solution}).
In particular, we see that bifurcations appear to occur at
\begin{equation}
\label{bif-emperic}
m = (2k-1)(1-n), \quad k \in \mathbb{N}.
\end{equation}
It is natural to conjecture that there is a countable number of bifurcations as $m$ increases beyond the values shown in the figure.

The present paper addresses bifurcations of self-similar solutions for reversing
and anti-reversing interfaces from the exact solution (\ref{exact-solution}).
We refer to the bifurcating solutions as the {\em secondary branches}, which emerge
from the primary branch. Since the existence of self-similar solutions is
defined by the ODE (\ref{ode}), the rest of this work focusses on analysis of
this equation only. Although our methods work for every $-1 \leq n < 1$, $m > 0$, and $m + n > 1$,
we will simplify many details by considering the case $n = 0$ and $m > 1$ only.

The paper is organized as follows. \S\ref{Kummer}
gives a quick review of properties of Kummer's differential equation.
\S\ref{proplinop} reports the results on linearization
of the ODE (\ref{ode}) at the exact solution (\ref{exact-solution}).
\S\ref{matched-asymptotics} describes a two-scale asymptotic method that allows us to obtain
the secondary branches near the bifurcation points by superposing the bifurcating mode
on the primary branch. In \S\ref{numerics} we compare the predictions of the asymptotic method
to numerical solutions and observe a good agreement between the two. Finally, \S\ref{conclusion} offers
physical interpretations of new self-similar solutions obtained in this work.

\section{Kummer's differential equation}
\label{Kummer}

Kummer's differential equation for the confluent hypergeometric functions is defined as follows:
\begin{equation}
\label{Kummer-ODE}
z \frac{d^2 w}{d z^2} + (b-z) \frac{dw}{dz} - a w = 0, \quad z \in \mathbb{R}^+,
\end{equation}
where $a,b \in \mathbb{R}$ are parameters. We refer the reader to either
Chapter 13 in \cite{AS} or Section 9.2 in \cite{GR} for a review of confluent hypergeometric functions.

The second-order ODE (\ref{Kummer-ODE}) has a regular singular point at $z = 0$ with two indices
$$
\sigma_1 = 0, \quad \sigma_2 = 1-b,
$$
and in what follows we consider $b > 1$ when $\sigma_2 < \sigma_1$.
If $b$ is not a non-positive integer (which is the case for our setting),
there exists a unique (up to a multiplicative constant)
bounded solution at $z = 0$ given by the following Kummer's function \cite{Kummer}
\begin{equation}
\label{Kummer-function}
M(z;a,b) := \sum_{n \in \mathbb{N}_0} \frac{a_n z^n}{b_n n!} =
1 + \frac{a}{b} \frac{z}{1!} + \frac{a(a+1)}{b(b+1)}\frac{z^2}{2!} + \frac{a(a+1)(a+2)}{b(b+1)(b+2)} \frac{z^3}{3!} + ...
\end{equation}
where $a_0 = 1$, $a_1 = a$, and $a_n = a (a+1) ... (a+n-1)$.

The other singular point of the second-order ODE (\ref{Kummer-ODE}) is $z = \infty$
and it is an irregular point with two linearly independent solutions
\begin{equation}
w_1(z) \sim z^{-a}, \quad w_2(z) \sim z^{a-b} e^{z} \quad \mbox{\rm as} \quad z \to \infty.
\end{equation}
There exists a unique (up to a multiplicative constant)
solution with the algebraic growth at infinity given by the following
Tricomi function \cite{Tricomi}
\begin{equation}
\label{Tricomi-function}
U(z;a,b) := \frac{\Gamma(1-b)}{\Gamma(1+a-b)} M(z;a,b) + \frac{\Gamma(b-1)}{\Gamma(a)} z^{1-b} M(z;1+a-b,2-b),
\end{equation}
such that the function $U(z;a,b)$ satisfies the asymptotic expansion at infinity (see 13.1.8 in \cite{AS}):
\begin{equation}
\label{Tricomi-infinity}
U(z;a,b) = z^{-a} \left[ 1 + \mathcal{O}(|z|^{-1}) \right] \quad \mbox{\rm as} \quad z \to +\infty.
\end{equation}

If $a$ is not a non-positive integer, then $U(z;a,b)$ is singular as $z \to 0$:
\begin{equation}
\label{Tricomi-zero}
U(z;a,b) = \frac{\Gamma(b-1)}{\Gamma(a)} z^{1-b} \left[ 1 + \mathcal{O}(z) \right] \quad \mbox{\rm as} \quad z \to 0,
\end{equation}
whereas $M(z;a,b)$ diverges at infinity (see 13.1.4 in \cite{AS}):
\begin{equation}
\label{Kummer-infinity}
M(z;a,b) = \frac{\Gamma(b)}{\Gamma(a)} z^{a-b} e^z \left[ 1 + \mathcal{O}(|z|^{-1}) \right] \quad \mbox{\rm as} \quad z \to +\infty.
\end{equation}

If $a$ is a non-positive integer, that is, if $a = -k$ with $k \in \mathbb{N}_0 := \{0,1,2,...\}$,
then Kummer's function $M(z;a,b)$ is truncated into a polynomial of degree $k$ so that $M(z;a,b)$ and $U(z;a,b)$
are linearly dependent. In \S\ref{proplinop}, we reduce the linearized equation
to Kummer's differential equation (\ref{Kummer-ODE}) and obtain the connection formulas
between $M(z;a,b)$ and $U(z;a,b)$ for a non-positive integer $a$.

\section{Linearization about the exact solution}
\label{proplinop}

The exact solution (\ref{exact-solution}) with the definition (\ref{aq_def}) can be written as
\begin{equation}
\label{exact-solution-zero}
H_Q(r) = r^{\frac{2}{m+1-n}},
\end{equation}
where $r := \xi/A_Q$. The differential equation (\ref{ode}) is linearized at the exact solution (\ref{exact-solution-zero})
by writing $H_- = H_Q + u$ and truncating at the linear terms in $u$. By doing so,
we obtain the homogeneous linear equation $Lu = 0$, where for $r \in \mathbb{R}^+$,
\begin{equation}
\label{linear-ode-x}
(L u)(r) := \frac{(m+1-n)^2}{2(m+1+n)} \frac{d^2}{d r^2} \left( r^{\frac{2m}{m+1-n}} u(r) \right)
- \frac{m+1-n}{2(1-n)} \, r \frac{d u}{d r}  + \frac{1}{1-n} u(r) - n r^{-\frac{2(1-n)}{m+1-n}} u(r).
\end{equation}
It is necessary to equip the differential equation $Lu = 0$ with suitable boundary conditions
at $r = 0$ and $r = \infty$. In order to consider the boundary conditions at $r = 0$,
let us use the following transformation of the independent variable
\begin{equation}
\label{transformation-x}
y := \frac{m+1-n}{1-n} r^{\frac{1-n}{m+1-n}},
\end{equation}
which allows us to rewrite the linear homogeneous equation $Lu = 0$ in the form
\begin{equation}
\label{linear-ode-y}
\frac{(m+1-n)^2}{2(m+1+n)} \left[ \frac{d^2 u}{d y^2} + \frac{3m}{(1-n) y} \frac{d u}{d y} + \frac{2(m+n) (m-n-1)}{(1-n)^2 y^2} u(y) \right]
- \frac{1}{2} \, y \frac{d u}{d y}  + \frac{1}{1-n} u(y) = 0,
\end{equation}
where we have used the same notation $u$ as a function of $y \in \mathbb{R}^+$.

Use of the Frobenius method (see Chapter 4 in \cite{Teschl}) reveals that $y = 0$
is a regular singular point of the differential equation (\ref{linear-ode-y})
with two indices $\sigma_{1,2}$ given by the indicial equation
\begin{equation}
\label{indices}
\sigma (\sigma - 1) + \frac{3 m}{1-n} \sigma + \frac{2 (m+n) (m-n-1)}{(1-n)^2} = 0 \quad \Rightarrow
\quad \sigma_1 = \frac{1+n-m}{1-n}, \quad \sigma_2 = -\frac{2(n+m)}{1-n}.
\end{equation}
We note that $\sigma_2 < \sigma_1$ for $m + 3n + 1 > 0$, which is satisfied if $m + n > 1$ and $1 + n \geq 0$.
Therefore, in what follows, we assume that $-1 \leq n < 1$.
As follows from the Frobenius method and the indicial equation (\ref{indices}), there exist two linearly independent solutions
of the differential equation (\ref{linear-ode-y}) with the following behaviours near $y = 0$:
\begin{equation}
\label{solution-1}
u_1(y) \sim y^{\frac{1+n-m}{1-n}} \quad \Rightarrow \quad
u_1(r) \sim r^{\frac{1+n-m}{m+1-n}}
\end{equation}
and
\begin{equation}
\label{solution-2}
u_2(y) \sim y^{-\frac{2(n+m)}{1-n}} \quad \Rightarrow  \quad
u_2(r) \sim r^{\frac{-2(n+m)}{m+1-n}}.
\end{equation}

In order to consider the boundary conditions at $r = \infty$ (or equivalently $y = \infty$),
let us use the following transformation of the dependent variable
\begin{equation}
\label{transformation-y}
u(y) = y^{-\frac{3m}{2(1-n)}} \exp \left( \frac{m+1+n}{4(m+1-n)^2} y^2 \right) v(y).
\end{equation}
On using (\ref{transformation-y}), the linear homogeneous equation (\ref{linear-ode-y}) can be rewritten in the form
\begin{equation}
\label{linear-ode-u}
- \frac{d^2 v}{d y^2} + \left[ \frac{m^2+2m+6mn+8n^2+8n}{4 (1-n)^2 y^2} - \frac{(3m+5-n)(m+1+n)}{2(1-n)(m+1-n)^2}
+ \frac{(m+1+n)^2 y^2}{4(m+1-n)^4} \right] v(y) = 0.
\end{equation}
The linear equation (\ref{linear-ode-u}) is well-known in quantum mechanics
as the stationary Schr\"{o}dinger equation for the multi-dimensional
harmonic oscillator \cite{quantum}. Owing to the harmonic confinement of the quantum oscillator,
there is only one linear independent solution of the differential equation (\ref{linear-ode-u})
that decays to zero as $y \to \infty$; the other solution grows rapidly as $y \to \infty$.
Using the Liouville-Green (JWKB) method, it can be shown that the two
linearly independent solutions behave at infinity as
\begin{equation}
\label{solution-1-inf}
v_1(y) \sim y^{\frac{3m+4}{2(1-n)}} \exp \left(-\frac{m+1+n}{4(m+1-n)^2} y^2 \right) \Rightarrow u_1(r) \sim r^{\frac{2}{m+1-n}}
\end{equation}
and
\begin{equation}
\label{solution-2-inf}
v_2(y) = y^{-\frac{3m+6-2n}{2(1-n)}} \exp \left( \frac{m+1+n}{4(m+1-n)^2} y^2\right) \Rightarrow
u_2(r) \sim r^{-\frac{3(m+1)-n}{m+1-n}} \exp \left( \frac{m+1+n}{2(1-n)^2} r^{\frac{2(1-n)}{m+1-n}}\right).
\end{equation}
The first solution matches the asymptotic behavior (\ref{ocean}), whereas the
second solution grows too fast and must be removed. Thus, in agreement with the invariant manifold result of Theorem 1.2 in \cite{FosterPel},
there is a unique (up to a normalizing constant) solution of the linear homogeneous equation $Lu = 0$,
where $L$ is given by (\ref{linear-ode-x}), which satisfies suitable behaviour at infinity.

The stationary Schr\"{o}dinger equation (\ref{linear-ode-u}) for the multi-dimensional
harmonic oscillator is solved in quantum mechanics at the admissible energy levels \cite{quantum}.
These energy levels correspond to the eigenfunctions $v$ of the linear equation (\ref{linear-ode-u}),
which belongs to $L^2(\mathbb{R}_+)$. For the function $u$ satisfying the homogeneous equation $Lu = 0$,
where $L$ is given by (\ref{linear-ode-x}), the admissible energy levels arise from the condition that
the algebraically growing solution (\ref{solution-1-inf}) at infinity
is connected to the slowest algebraically growing solution (\ref{solution-1}) at zero.

Compared to this conventional treatment of the stationary Schr\"{o}dinger equation (\ref{linear-ode-u}),
we will need to clarify the role of both algebraically growing solutions (\ref{solution-1}) and (\ref{solution-2})
at zero in the context of the self-similar solutions of the ODE (\ref{ode}).
Therefore, we need a general solution of the stationary Schr\"{o}dinger equation (\ref{linear-ode-u})
satisfying (\ref{solution-1-inf}), which is the only behaviour allowed there
for the self-similar solutions of the ODE (\ref{ode}).

In order to reduce the stationary Schr\"{o}dinger equation (\ref{linear-ode-u}) to the Kummer
differential equation (\ref{Kummer-ODE}), we transform both the dependent and independent variables as follows:
\begin{equation}
\label{transformation-z}
z := \frac{(m+1+n)}{2(m+1-n)^2} y^2, \quad v(y) := z^{\frac{m+2+2n}{4(1-n)}} \exp \left( -\frac{z}{2} \right) w(z).
\end{equation}
On doing so, the linear homogeneous equation (\ref{linear-ode-u}) can be rewritten in the form
\begin{equation}
\label{linear-ode-w}
z \frac{d^2 w}{d z^2} + \left[ \frac{m+3+n}{2(1-n)} - z \right] \frac{d w}{dz} + \frac{m+1-n}{2(1-n)} w(z) = 0,
\end{equation}
which coincides with Kummer's equation (\ref{Kummer-ODE}) for
\begin{equation}
\label{a-b}
a := -\frac{m+1-n}{2(1-n)}, \quad b := \frac{m+3+n}{2(1-n)}.
\end{equation}
Kummer's function $M(z;a,b)$ defined by (\ref{Kummer-function})
behaves near zero like the first slowest growing solution (\ref{solution-1})
after the transformations (\ref{transformation-x}), (\ref{transformation-y}), and (\ref{transformation-z})
have been used. If $a$ is not a non-positive integer, then $M(z;a,b)$
satisfies (\ref{Kummer-infinity}), which corresponds to the second, fastest growing, solution (\ref{solution-2-inf}) at infinity.
If $a$ is a non-positive integer, then $M(z;a,b)$ is truncated into a polynomial, which corresponds
to the first, slowest growing, solution (\ref{solution-1-inf}). This happens for $a = a_k$ and $b = b_k$, where
\begin{equation}
\label{bifurcation-a-b}
a_k := -k, \quad b_k := k + \frac{1+n}{1-n}, \quad k \in \mathbb{N}_0 := \{0,1,2,...\},
\end{equation}
in which case,
\begin{equation}
\label{bifurcation-value}
m = m_k := (2k -1) (1-n), \quad k \in \mathbb{N}_0.
\end{equation}
Note that the bifurcation points given by (\ref{bifurcation-value}) coincide with
(\ref{bif-emperic}), except for the additional point $m_0 = n-1$.
Since we are only interested in values of $m > 0$ and $n < 1$ the bifurcation point
at $m_0 = n-1 < 0$ can be ignored.

Let us state explicitly the polynomials arising at the first four bifurcation points:
\begin{eqnarray}
\label{polynomial-1}
k = 1 &: & M\left(z;a_1,b_1\right) = 1 - \frac{1-n}{2} z, \\
\label{polynomial-2}
k = 2 &: & M\left(z;a_2,b_2\right) = 1 - \frac{2(1-n)}{3-n} z + \frac{(1-n)^2}{2(3-n)(2-n)} z^2, \\
\nonumber
k = 3 &: & M\left(z;a_3,b_3\right) = 1 - \frac{3(1-n)}{2(2-n)} z + \frac{3(1-n)^2}{2(2-n)(5-3n)} z^2 \\
\label{polynomial-3}
&\phantom{t}& \phantom{text} - \frac{(1-n)^3}{4 (2-n)(5-3n)(3-2n)} z^3, \\
\nonumber
k = 4 &: &  M(z;a_4,b_4) = 1 - \frac{4(1-n)}{5-3n} z + \frac{13(1-n)^2}{(5-3n)(3-2n)} z^2 \\
&\phantom{t}& \phantom{text}  - \frac{2(1-n)^3}{(5-3n)(3-2n)(7-5n)} z^3 + \frac{(1-n)^4}{4(5-3n)(3-2n)(7-5n)(4-3n)} z^4.
\label{polynomial-4}
\end{eqnarray}

For every $a$ and $b$, Tricomi's function $U(z;a,b)$ defined by (\ref{Tricomi-function})
is the only solution of the Kummer's differential equation (\ref{Kummer-ODE})
satisfying the asymptotic behaviour (\ref{Tricomi-infinity}), which
corresponds to the slowest growing solution (\ref{solution-1-inf}) at infinity,
after the transformations (\ref{transformation-x}), (\ref{transformation-y}), and (\ref{transformation-z})
have been used. If $a$ is not a non-positive integer, then $U(z;a,b)$
satisfies (\ref{Tricomi-zero}), which corresponds to the fastest growing solution (\ref{solution-2}) near zero.

The projection of $U(z;a,b)$ to the fastest growing solution (\ref{solution-2}) near zero
is defined by taking the limit for $b > 1$ (which is satisfied in our case for $m + 3n + 1 > 0$):
\begin{eqnarray}
\nonumber
B(m) & := & \lim_{z \to 0} z^{b-1} U(z;a,b) = \frac{\Gamma(b-1)}{\Gamma(a)}
= \frac{\Gamma\left(\frac{m+1+3n}{2(1-n)}\right)}{\Gamma\left(-\frac{m+1-n}{2(1-n)}\right)}\\
& = & \pi^{-1} \Gamma\left(\frac{m+1+3n}{2(1-n)}\right) \Gamma\left(\frac{m+3-3n}{2(1-n)}\right) \sin \left( \frac{\pi(m+3-3n)}{2(1-n)} \right),
\label{constant-B-exact}
\end{eqnarray}
where we have used the following continuation property of the Gamma function (see 8.334 in \cite{GR})
\begin{equation}
\label{continuation-gamma}
\Gamma(x) \Gamma(1-x) = \frac{\pi}{\sin \pi x}.
\end{equation}
We note that $B(m_k) = 0$ at the bifurcation point (\ref{bifurcation-value}) and
\begin{equation}
\label{constant-B}
B'(m_k) = \frac{1}{2 (1-n)} (-1)^{k+1} \Gamma\left(k + \frac{2n}{1-n}\right) \Gamma(k+1).
\end{equation}

It is more difficult to compute the projection of $U(z;a,b)$ to the slowest growing solution (\ref{solution-1})
near zero at the bifurcation point $m = m_k$. The first term in (\ref{Tricomi-function}) gives the projection
to the slowest growing solution (\ref{solution-1}) which is characterized by the quantity
\begin{eqnarray}
\label{def-C-m-k}
C(m_k) := \lim_{m \to m_k}
\frac{\Gamma\left(-\frac{m+1+3n}{2(1-n)}\right)}{\Gamma\left(-\frac{m+1+n}{1-n}\right)}
= \lim_{m \to m_k} \frac{\Gamma\left(\frac{m+2}{1-n}\right)}{\Gamma\left(\frac{m+3+n}{2(1-n)}\right)}
\frac{\sin \left( \frac{\pi (m+2)}{1-n} \right)}{\sin \left( \frac{\pi (m+3+n)}{2(1-n)} \right)}.
\end{eqnarray}
Let us define
\begin{equation}
\label{constant-p}
p := \frac{2n}{1-n}.
\end{equation}
Then, the limit $m \to m_k$ in (\ref{def-C-m-k}) yields the following explicit expression
\begin{eqnarray}
C(m_k) = \left\{ \begin{array}{ll} (-1)^k \frac{\Gamma\left(2k+1+p\right)}{\Gamma\left(k+1+p\right)}, \quad & p \notin \mathbb{Z} \\
2 (-1)^k \frac{(2k+p)!}{(k+p)!}, \quad & p \in \mathbb{Z}, \;\; k+p \in \mathbb{N},\end{array} \right.
\label{constant-C-exact}
\end{eqnarray}
where the continuation formula (\ref{continuation-gamma}) has been used as well as the elementary
property $\Gamma(k+1) = k!$ for a positive integer $k$. The second
term in (\ref{Tricomi-function}) does not give a projection to the slowest growing solution (\ref{solution-1})
if $b_k-1$ is not an integer, that is, when $p \notin \mathbb{Z}$. On the other hand,
when $p \in \mathbb{Z}$, we have
\begin{equation}
z^{1-b_k} M(z;1+a_k-b_k,2-b_k) = z^{-k-p} M(z;-2k-p,1-k-p)
\end{equation}
and since $-2k-p < 1-k-p$, the function above is not defined if $1-k-p$ is a non-positive integer.
In order to resolve the singularity, we note the limit 9.214 in \cite{GR} for $k+p \in \mathbb{N}$:
\begin{equation}
\label{9-214}
\lim_{b \to b_k} \frac{M(z;1+a-b,2-b)}{\Gamma(2-b)} = \left( \begin{array}{c} a-1 \\ k+p \end{array} \right)
z^{k+p} M(z;a,b_k).
\end{equation}
Therefore, we obtain from (\ref{Tricomi-function}) and (\ref{9-214}) for $k + p \in \mathbb{N}$:
\begin{eqnarray}
\nonumber
D(m_k) & := & \lim_{z \to 0} \lim_{m \to m_k} \frac{\Gamma(b-1) \Gamma(2-b)}{\Gamma(a)} z^{1-b} \frac{M(z;1+a-b,2-b)}{\Gamma(2-b)}  \\
& = & \left( \begin{array}{c} -1-k \\ k+p \end{array} \right)   \lim_{m \to m_k} \frac{\Gamma(1-a) \sin \pi(1-a)}{\sin \pi(b-1)},
\end{eqnarray}
where the continuation formula (\ref{continuation-gamma}) has been used. This yields
\begin{eqnarray}
\nonumber
D(m_k) & = & (-1)^{k+p} \frac{(2k+p)!}{k! (k+p)!}
\lim_{m \to m_k} \frac{\Gamma\left(\frac{m-3(1-n)}{2(1-n)}\right) \sin \left( \pi \left(\frac{m-3(1-n)}{2(1-n)}\right) \right)}{\sin \left( \pi \left(\frac{m+3n+1}{2(1-n)}\right) \right)}\\
& = & (-1)^{k+1} \frac{(2k+p)!}{(k+p)!}.
\label{constant-D-exact}
\end{eqnarray}
Thus, for $p \in \mathbb{Z}$, we define
\begin{equation}
\label{constant-E-exact}
E(m_k) := C(m_k) + D(m_k) = (-1)^k \frac{(2k+p)!}{(k+p)!}.
\end{equation}
Note that this expression is a limit of $C(m_k)$ in the first line of (\ref{constant-C-exact}) when a non-integer $p$
approaches an integer value.

In \S \ref{matched-asymptotics}, the asymptotic formulas (\ref{bifurcation-a-b}), (\ref{constant-B}), and
(\ref{constant-E-exact}) are incorporated into the construction of the self-similar solution
to the ODE (\ref{ode}) near the bifurcation point $m = m_k$, $k \in \mathbb{N}$.

\section{\label{matched-asymptotics}Two-scale asymptotic method for bifurcating solutions}

Here we consider the differential equation (\ref{ode}) with $n  = 0$, the latter simplification
is made purely to reduce what would otherwise be cumbersomely large equations.
However, we do note that the cases $n \neq 0$ with $-1 \leq n < 1$ can be included in
our asymptotic analysis. After the subscript is dropped, the second-order
ODE (\ref{ode}) with $n  = 0$ is written in the form:
\begin{equation}
\label{ode-before}
\frac{d}{d \xi} \left( H^m \frac{d H}{d \xi} \right)
- \frac{m+1}{2} \, \xi \frac{d H}{d \xi} = 1 - H.
\end{equation}
We are looking for the monotonically increasing solution on $[\hat{\xi},\infty)$ with
\begin{equation}
\label{near-boundary-again}
\left\{ \begin{array}{l} H(\xi) \to 0, \\
H^m(\xi) H'(\xi) \to 0, \end{array} \right. \quad \mbox{\rm as} \quad \xi \to \hat{\xi}
\end{equation}
and
\begin{equation}
\label{far-boundary-again}
H(\xi) \sim \left(\frac{\xi}{A}\right)^{\frac{2}{m+1}} \quad \mbox{\rm as} \quad \xi \to \infty,
\end{equation}
for some $A > 0$, see (\ref{ocean}), (\ref{onebeh}), and (\ref{genes}).

In \S \ref{sec-inner}, we consider suitable solutions near $\xi \gtrsim \hat{\xi}$ for small $\hat{\xi}$.
In \S \ref{sec-outer}, we expand solutions near the exact solution
\begin{equation}
\label{exact-again-section}
H_Q(\xi) = \left( \frac{\xi}{A_Q} \right)^{\frac{2}{m+1}}, \quad A_Q = \frac{\sqrt{2}}{\sqrt{m+1}},
\end{equation}
which corresponds to the case $\hat{\xi} = 0$ for $n = 0$. Matching conditions
between the two formal asymptotic expansions are considered in \S \ref{sec-matching},
where small $\hat{\xi}$ and $A-A_Q$ are uniquely defined in terms of $m - m_k$,
where $m_k = (2k-1)$, $k \in \mathbb{N}$ is the bifurcation point for $n = 0$.
Particular computations for $k = 1, 2, 3, 4$ are given as examples of these bifurcations.

\subsection{Inner scale}
\label{sec-inner}

In order to study the behaviour of solutions both in the near field (near the interface at $\xi = \hat{\xi}$), and in
the asymptotic limit $\hat{\xi} \to 0$ (that is, close to the bifurcation value $m = m_k$, $k \in \mathbb{N}$),
we use the scaling transformation
\begin{equation}
\label{scaling}
\xi = \hat{\xi} + |\hat{\xi}|^{\frac{m+1}{m-1}} \eta, \quad H(\xi) = |\hat{\xi}|^{\frac{2}{m-1}} \mathcal{H}(\eta),
\end{equation}
where $\mathcal{H}$ satisfies the second-order ODE
\begin{equation}
\label{ode-before-scaled}
\frac{d}{d \eta} \left( \mathcal{H}^m \frac{d \mathcal{H}}{d \eta} \right)
= 1 + \frac{m+1}{2} \sigma \frac{d \mathcal{H}}{d \eta} + |\hat{\xi}|^{\frac{2}{m-1}} \left( \frac{m+1}{2} \eta \frac{d \mathcal{H}}{d \eta} - \mathcal{H} \right),
\end{equation}
where $\sigma = {\rm sign}(\hat{\xi})$. In the limit $\hat{\xi} \to 0$,
the second-order ODE (\ref{ode-before-scaled})
is truncated to the autonomous equation, which can be integrated once
with the boundary conditions obtained from (\ref{near-boundary-again}):
\begin{equation}
\left\{ \begin{array}{l} \mathcal{H}(\eta) \to 0, \\
\mathcal{H}^m(\eta) \mathcal{H}'(\eta) \to 0, \end{array} \right.  \quad \mbox{\rm as} \quad \eta \to 0.
\end{equation}
After truncation and integration, the resulting equation is
\begin{equation}
\label{first-order}
\mathcal{H}_0^m \frac{d \mathcal{H}_0}{d \eta} = \eta + \frac{m+1}{2} \sigma \mathcal{H}_0,
\end{equation}
where $\mathcal{H}_0$ denotes the leading order of the solution $\mathcal{H}$ after truncation.
The first-order non-autonomous equation (\ref{first-order}) is equivalent to
the following planar dynamical system
\begin{equation}
\left\{ \begin{array}{l} \dot{\eta} = \mathcal{H}_0^m, \\
\dot{\mathcal{H}}_0 = \eta + \frac{m+1}{2} \sigma \mathcal{H}_0, \end{array} \right.
\label{dyn-sys}
\end{equation}
where the dot denotes a derivative with respect to the `time' variable $\tau$.
The point $(\eta,\mathcal{H}_0) = (0,0)$ is the only equilibrium point of
the planar system (\ref{dyn-sys}). If $m > 1$ (since $m + n > 1$ and $n = 0$),
the equilibrium point $(0,0)$ is located at the intersection of
a center curve tangential to the straight line
\begin{equation}
\label{center-manifold}
E^c(0,0) = \left\{ \eta = -\frac{m+1}{2} \sigma \mathcal{H}_0, \quad \mathcal{H}_0 \in \mathbb{R} \right\}
\end{equation}
and an unstable (stable) curve for $\hat{\xi} > 0$ ($\hat{\xi} < 0$), which is
tangential to the $\mathcal{H}_0$-axis.

We are only interested in constructing a trajectory of the dynamical system (\ref{dyn-sys})
in the first quadrant where $\mathcal{H}_0 > 0$ and $\eta > 0$. If $\hat{\xi} > 0$ ($\sigma = +1$), the tangent line $E^c(0,0)$ in (\ref{center-manifold})
to the center curve is not located in the first quadrant. Therefore, there is a unique
trajectory of the dynamical system (\ref{dyn-sys}) that departs from
$(0,0)$ in the first quadrant along the unstable curve and satisfies
the exponential growth
$$
\mathcal{H}_0(\tau) \sim h_0 \exp \left( \frac{m+1}{2} \tau \right), \quad
\eta(\tau) \sim \frac{2 h_0^m}{m(m+1)} \exp \left( \frac{m (m+1)}{2}\tau \right) \quad \mbox{\rm as} \quad \tau \to -\infty,
$$
where $h_0 > 0$ is an arbitrary constant. This yields the asymptotic expression
\begin{equation}
\label{leading-zero}
\mathcal{H}_0(\eta) \sim \left( \frac{m(m+1)}{2} \eta \right)^{\frac{1}{m}} \quad \mbox{\rm as} \quad \eta \to 0,
\end{equation}
which coincides with the asymptotic behaviour (\ref{onebeh}) for $n = 0$ in near-field
after the change of variables (\ref{scaling}).

If $\hat{\xi} < 0$ ($\sigma = -1$), the tangent line $E^c(0,0)$ in (\ref{center-manifold})
to the center curve is now located in the first quadrant. Since the other
invariant curve is stable, there is a unique trajectory of the dynamical system (\ref{dyn-sys}) that departs from
$(0,0)$ in the first quadrant along the center curve. The trajectory satisfies
\begin{equation}
\label{leading-zero-minus}
\mathcal{H}_0(\eta) \sim \frac{2}{m+1} \eta \quad \mbox{\rm as} \quad \eta \to 0,
\end{equation}
which coincides with the asymptotic behaviour (\ref{genes}) for $n = 0$ in near-field
after the change of variables (\ref{scaling}).

If $\hat{\xi} > 0$ ($\sigma = +1$), it follows from the first-order equation (\ref{first-order}) that if a solution
originates from the point $(\eta,\mathcal{H}_0) = (0,0)$ in the first quadrant, then
$\mathcal{H}_0$ is an monotonically increasing function of $\eta$ with no stopping points.
If $\hat{\xi} < 0$ ($\sigma = -1$), the same can be concluded by a contradiction. Suppose there is a finite `time'
$\tau_0$ and a finite $\mathcal{H}_0(\tau_0) > 0$ such that $\dot{\mathcal{H}_0}(\tau_0) = 0$. Then, it
follows from the system (\ref{dyn-sys}) that $\ddot{\mathcal{H}_0}(\tau_0) = \dot{\eta}(\tau_0) =
\mathcal{H}_0^m(\tau_0) > 0$, so that $\tau_0$ is a minimum of $\mathcal{H}_0$ as a function of $\tau_0$.
However, this contradicts to the fact that $\mathcal{H}_0$ was an increasing function of $\tau$ for $\tau < \tau_0$.

Thus, for both $\hat{\xi} > 0$ and $\hat{\xi} < 0$, the unique solution of the first-order equation (\ref{first-order})
reaches infinity and since the right-hand side of the second equation in system (\ref{dyn-sys})
is linear in $\mathcal{H}_0$, the solution cannot reach infinity in a finite $\eta$.
Therefore, the unique solution satisfies $\mathcal{H}_0 \to \infty$
as $\eta \to \infty$. In order to derive the asymptotic behavior of the solution
near infinity, we rewrite (\ref{first-order}) following another integration with respect to $\eta$. We have
\begin{equation}
\label{integral-eq}
\frac{1}{m+1} \mathcal{H}_0^{m+1} = \frac{1}{2} \eta^2 + \frac{m+1}{2} \sigma \int_0^{\eta} \mathcal{H}_0(\eta') d \eta'.
\end{equation}
It is now easy to obtain the asymptotic behaviour of $\mathcal{H}_0$ as $\eta \to \infty$
by iteration. We find that
\begin{equation}
\label{leading-infinity}
\mathcal{H}_0(\eta) \sim \left( \frac{m+1}{2} \eta^2 \right)^{\frac{1}{m+1}} \quad \mbox{\rm as} \quad \eta \to \infty.
\end{equation}
This behaviour coincides with the asymptotic behaviour (\ref{far-boundary-again})
with $A = A_Q$ and $n = 0$ in the far-field after the change of variables (\ref{scaling}).

Summarizing, we have proved the existence of a unique solution to the truncated first-order equation (\ref{first-order})
which satisfies the leading-order asymptotic expansions (\ref{leading-zero}) or (\ref{leading-zero-minus}) at zero
and the leading-order asymptotic expansion (\ref{leading-infinity})
at infinity. The behaviour at infinity for the full solution $\mathcal{H}(\eta)$ is subject to the remainder terms
proportional to $|\hat{\xi}|^{2/(m-1)}$ in the second-order equation (\ref{ode-before-scaled}).

We note that the expansion near infinity with the leading-order term in (\ref{leading-infinity})
is only understood in the asymptotic sense. It was proved in \cite{FosterPel} that the trajectory
of the differential equation (\ref{ode-before}) that originates at $H(\hat{\xi}) = 0$ and extends to $H(\xi) > 0$
for $\xi > \hat{\xi}$ does not generally reach infinity but turns back towards smaller values of $H$.
It is only for special values of $\hat{\xi}$, that this trajectory reaches infinity to give curves
on the solution diagram of Figure \ref{all-sols-negative-t}. In order to find these special values
of $\hat{\xi}$, in the limit of small $\hat{\xi}$, \emph{i.e.}, near the bifurcations,
we need to construct the outer expansion and to deduce the asymptotic matching conditions on the two-scale expansions.

\subsection{Outer scale}
\label{sec-outer}

Let us consider solutions of the ODE (\ref{ode-before}) in the neighborhood
of the exact solution (\ref{exact-again-section}), which is defined for every $\xi > 0$.
To do so, we use the following regular asymptotic expansion:
\begin{equation}
\label{outer}
H(\xi) = H_Q(r) + \alpha u_1(r) + \alpha^2 u_2(r) + \mathcal{O}(\alpha^3), \quad r := \frac{\xi}{A_Q},
\end{equation}
where $\alpha \in \mathbb{R}$ is the small parameter in the formal expansion and
the correction terms $\{u_1, u_2, \cdots\}$ are to be defined recursively subject to appropriate
boundary conditions.

Inserting the expansion (\ref{outer}) into (\ref{ode-before})
and balancing terms at $\mathcal{O}(\alpha)$, we obtain the homogeneous linear equation $L u_1 = 0$,
where the linear operator $L$ is given by (\ref{linear-ode-x}) with $n = 0$.
We have proved in \S\ref{proplinop} that there is
only one solution of the homogeneous equation $L u_1 = 0$
(up to a multiplicative factor given by $\alpha$), which satisfies the slowest growing
behaviour (\ref{solution-1-inf}) as $r \to \infty$. This solution is
given by Tricomi's function $U(z;a,b)$ in (\ref{Tricomi-function}).
After employing the transformations (\ref{transformation-x}), (\ref{transformation-y}), (\ref{transformation-z}),
and (\ref{a-b}) with $n = 0$, we can define $u_1$ in terms of $r$ as
\begin{equation}
\label{solution-u-1}
u_1(r) = r^{\frac{1-m}{1+m}} U\left(\frac{m+1}{2} r^{\frac{2}{m+1}}; -\frac{m+1}{2},\frac{m+3}{2}\right).
\end{equation}

Proceeding to balance terms at $\mathcal{O}(\alpha^2)$, we obtain the linear inhomogeneous equation
\begin{equation}
\label{inhomogeneous-eq}
L u_2 = R_2 := -\frac{m(m+1)}{4} \frac{d^2}{dr^2} \left[ r^{\frac{2(m-1)}{m+1}} u_1^2 \right].
\end{equation}
Owing to the asymptotic behaviour (\ref{solution-1-inf}) for $u_1(r)$ as $r \to \infty$,
$R_2(r)$ is bounded as $r \to \infty$ and converges to a constant. Similarly, we observe
that
$$
L r^{\frac{2}{m+1}} = (m+1),
$$
therefore, there exists a solution of the inhomogeneous equation (\ref{inhomogeneous-eq})
satisfying the same asymptotic behaviour (\ref{solution-1-inf}) at infinity. This solution is defined
up to the choice of the homogeneous solution proportional to $u_1$ given by (\ref{solution-u-1}).
Altering this choice of the homogeneous solution simply corresponds to redefining the small parameter
$\alpha$ in the expansion (\ref{outer}). Therefore, without loss of generality, the homogeneous solution
can be removed from the definition of $u_2(r)$, which then becomes uniquely defined.

Now, let us consider the behavior of solutions of the inhomogeneous equation (\ref{inhomogeneous-eq})
near $r = 0$. From (\ref{constant-B-exact}) with $n = 0$, we know that
\begin{equation}
U(z;a,b) \sim B(m) z^{-\frac{m+1}{2}}  \quad \mbox{\rm as} \quad z \to 0.
\end{equation}
From (\ref{solution-u-1}), this yields the asymptotic behaviour
\begin{equation}
\label{dominant-u-1}
u_1(r) \sim B(m) 2^{\frac{m+1}{2}} (m+1)^{-\frac{m+1}{2}} r^{-\frac{2m}{m+1}} \quad \mbox{\rm as} \quad r \to 0.
\end{equation}
If $B(m) \neq 0$, then $R_2(r) \sim r^{-4}$ as $r \to 0$, so that the
linear inhomogeneous equation (\ref{inhomogeneous-eq}) produces the
solution (up to a multiplicative factor)
\begin{equation}
u_2(r) \sim B(m) r^{-\frac{4m+2}{m+1}}, \quad \mbox{\rm as} \quad r \to 0.
\end{equation}
If $B(m) \neq 0$, the outer expansion (\ref{outer}) becomes singular with the fastest growth as $r \to 0$,
which cannot be matched with the inner expansion obtained from (\ref{scaling}) and (\ref{leading-infinity}).
However, we show in \S \ref{sec-matching} that the inner and outer expansions can be matched together
near $m = m_k$ for some $k \in \mathbb{N}$ since $B(m_k) = 0$.

Recall that for $n = 0$, we have $p = 0$ in (\ref{constant-p}), so that
both (\ref{constant-C-exact}) and (\ref{constant-D-exact}) are used
to yield (\ref{constant-E-exact}) for $m = m_k$, $k \in \mathbb{N}$. By using (\ref{solution-u-1}), we find
\begin{equation}
\label{dominant-u-C-n}
m = m_k : \quad u_1(r) = E(m_k)
r^{\frac{1-m}{1+m}} M\left(\frac{m+1}{2} r^{\frac{2}{m+1}};-k,k+1\right),
\end{equation}
where $E(m_k)$ is defined by (\ref{constant-E-exact}).
By using (\ref{inhomogeneous-eq}), we obtain the inhomogeneous term of the linear equation in the following form,
\begin{equation}
\label{dominant-R-2-n}
R_2(r) = -\frac{m(m+1)}{4} E(m_k)^2 \frac{d^2}{d r^2} \left[ M^2\left(\frac{m+1}{2} r^{\frac{2}{m+1}};-k,k+1\right) \right].
\end{equation}
Since $M(z;-k,k+1)$ is a polynomial of degree $k$ in $z$ given by (\ref{Kummer-function}),
the source term contains powers of $r^{2(-m+\ell)/(m+1)}$ for the integer $\ell$ counted from
$0$ to $m = m_k = 2k-1$ with the missing factor at $\ell_k = (m-1)/2 = k-1$. The dominant term
$r^{-2m/(m+1)}$ in $R_2(r)$ generates the same term in the solution $u_2(r)$ since
$$
L r^{-\frac{2m}{m+1}} = (m+1) r^{-\frac{2m}{m+1}}.
$$
Therefore, in the case $B(m_k) = 0$, we obtain
\begin{equation}
\label{constant-D}
u_2(r) \sim F(m_k) r^{-\frac{2m}{m+1}} \quad \mbox{\rm as} \quad r \to 0,
\end{equation}
where $F(m_k)$ is computed from a linear algebraic system. Note that the dominant term
in (\ref{constant-D}) is comparable with the dominant term (\ref{dominant-u-1}) in the solution $u_1$ for $m \neq m_k$.

\subsection{\label{matching_sec}Matching conditions}
\label{sec-matching}

Here we match the two (inner and outer) asymptotic regions together. Using the scaling transformation (\ref{scaling}) and the leading-order behaviour (\ref{leading-infinity}), we obtain the dominant term of the inner expansion as follows
\begin{equation}
\label{matching-inner}
H(\xi) \sim \left( \frac{\xi - \hat{\xi}}{A_Q} \right)^{\frac{2}{m+1}}, \quad \mbox{\rm as} \quad
\frac{\xi - \hat{\xi}}{|\hat{\xi}|^{\frac{m+1}{m-1}}} \to \infty \quad \mbox{\rm and} \quad \hat{\xi} \to 0.
\end{equation}
Expanding as $\hat{\xi} \to 0$ and using $r$ as in (\ref{outer}), we obtain
\begin{equation}
\label{matching-inner-expanded}
H(\xi) \sim r^{\frac{2}{m+1}} - \frac{2 \hat{\xi} }{(m+1) A_Q} r^{\frac{1-m}{1+m}}
+ \frac{(1-m) \hat{\xi}^2}{2(m+1)} r^{-\frac{2m}{m+1}}, \quad \mbox{\rm as} \quad
\frac{\xi - \hat{\xi}}{|\hat{\xi}|^{\frac{m+1}{m-1}}} \to \infty \quad \mbox{\rm and} \quad \hat{\xi} \to 0.
\end{equation}

We can see that the first two correction terms in (\ref{matching-inner-expanded})
occur also in the first two perturbation terms of the outer expansion (\ref{outer}) seen in
(\ref{dominant-u-1}), (\ref{dominant-u-C-n}), and (\ref{constant-D}). This suggests that two constraints should arise from the matching process. The first constraint on the slowest
growing term defines the parameter $\alpha$ in terms of $\hat{\xi}$:
\begin{equation}
\label{matching-condition-1}
- \frac{2 \hat{\xi} }{(m+1) A_Q}  = \alpha E(m_k) + \mathcal{O}(\alpha (m-m_k),\alpha^2).
\end{equation}
The second constraint on the fastest growing term defines $m-m_k$ in terms of either $\alpha$ or $\hat{\xi}$:
\begin{equation}
\label{matching-condition-2}
\frac{(1-m) \hat{\xi}^2}{2(m+1)}  = \alpha B'(m_k) (m-m_k) 2^{\frac{m+1}{2}} (m+1)^{-\frac{m+1}{2}} + \alpha^2 F(m_k)
+ \mathcal{O}(\alpha (m-m_k)^2, \alpha^2 (m-m_k), \alpha^3).
\end{equation}
After $\alpha$ is eliminated from the system (\ref{matching-condition-1}) and (\ref{matching-condition-2}),
we obtain an asymptotic approximation of the solution curve in the $(\hat{\xi},m)$-plane.

On the other hand, in the limit $r \to \infty$, we compare the outer asymptotic expansion (\ref{outer})
with the asymptotic behaviour (\ref{far-boundary-again}) at infinity,
where $A$ is a parameter. From (\ref{Tricomi-infinity}) and (\ref{solution-u-1}), we obtain
a constraint that defines the parameter $\alpha$ in terms of $A - A_Q$:
\begin{equation}
\label{matching-condition-3}
\left(\frac{A_Q}{A} \right)^{\frac{2}{m+1}}  = 1 + \alpha (m+1)^{\frac{m+1}{2}} 2^{-\frac{m+1}{2}} + \mathcal{O}(\alpha^2).
\end{equation}
Equation (\ref{matching-condition-3}) yields the asymptotic approximation of the solution curve
in the $(A,\hat{\xi})$-plane in view of equation (\ref{matching-condition-1}) or
in the $(A,m)$-plane in view of the dependence of $\hat{\xi}$ versus $m$.

The sign-alternation of $E(m_k)$ over $k \in \mathbb{N}$ given by (\ref{constant-E-exact})
yields by virtue of (\ref{matching-condition-1}) and (\ref{matching-condition-3})
the sign alternation of the dependence $(A-A_Q)$ versus $\hat{\xi}$ near the bifurcation point
where $A = A_Q$ and $\hat{\xi} = 0$. This fact explains why the location of the red and blue curves
bifurcating above and below the black curve on figure \ref{all-sols-negative-t}
alternates between the two adjacent bifurcation points.

In order to compare our analytical and numerical approaches, let us now compute
the asymptotic dependencies near the first four bifurcation points explicitly.

\subsubsection{\label{localm1}Behaviour local to $m = 1$:}
\label{Section3-3-1}

At $m = 1$, there exists a one-parameter family of exact solutions to the differential equation (\ref{ode-before})
given by
\begin{equation}
\label{mEq1}
H(\xi) = a (\xi - \hat{\xi}), \quad \hat{\xi} = \frac{a^2-1}{a}, \quad a \in \mathbb{R}.
\end{equation}
We show that the matching conditions (\ref{matching-condition-1}), (\ref{matching-condition-2}), and (\ref{matching-condition-3})
recover the exact solution (\ref{mEq1}). This implies that no new solution branches bifurcate near $m = 1$.

On setting $k=1$ and $n = 0$ in (\ref{constant-B}) and (\ref{constant-E-exact}), we obtain
$B'(1) = 1/2$ and $E(1) = -2$.
Since $A_Q = 1$, the matching conditions (\ref{matching-condition-1}) and (\ref{matching-condition-3}) tell us that
\begin{equation}
\label{m1zero}
\hat{\xi} = 2 \alpha + \mathcal{O}(\alpha^2), \quad \frac{1}{A} = 1 + \alpha + \mathcal{O}(\alpha^2).
\end{equation}
From (\ref{polynomial-1}) with $n = 0$, (\ref{dominant-u-C-n}), and (\ref{dominant-R-2-n}), we obtain
$u_1(r) = r - 2$ and $R_2(r) = -1$. Since $L r^0 = 1$ if $m = 1$ and $n = 0$,
there is a unique solution $u_2(r) = -1$
of the linear equation (\ref{inhomogeneous-eq}), from which we obtain $F(1) = 0$ from
(\ref{constant-D}). The matching condition (\ref{matching-condition-2}) yields
\begin{equation}
\label{m1inf}
m - 1 = \mathcal{O}(\alpha^2).
\end{equation}

Although the approximation (\ref{m1inf}) may imply that $m \neq 1$ for $\alpha \neq 0$ (or $\hat{\xi} \neq 0$),
let us observe the correspondence between the asymptotic solution (\ref{outer}) with $\xi = r$, given explicitly by
\begin{equation}
\label{m1expansion}
H(\xi) = \xi + \alpha (\xi - 2) - \alpha^2 + \mathcal{O}(\alpha^3),
\end{equation}
and the exact solution (\ref{mEq1}) with $a = 1 + \alpha$, which yields (\ref{m1expansion}) with
the $\mathcal{O}(\alpha^3)$ remainder term being equal to zero. From
(\ref{far-boundary-again}), (\ref{mEq1}), and (\ref{m1expansion}), we obtain
$$
\hat{\xi} = \alpha \frac{2 + \alpha}{1 + \alpha}, \quad A = \frac{1}{1+\alpha},
$$
which shows that the remainder terms in the second formula (\ref{m1zero}) and
in (\ref{m1inf}) are identically zero.

\subsubsection{\label{localm3}Behaviour local to $m = 3$:}

On setting $k=2$ and $n = 0$ in (\ref{constant-B}) and (\ref{constant-E-exact}), we obtain
$B'(3) = -1$ and $E(3) = 12$. Since $A_Q = 1/\sqrt{2}$, we obtain from (\ref{matching-condition-1}) and (\ref{matching-condition-3}):
$$
\hat{\xi} = - 12 \sqrt{2} \alpha + \mathcal{O}(\alpha^2), \quad \frac{A_Q}{A} = 1 + 8 \alpha + \mathcal{O}(\alpha^2),
$$
that is,
\begin{equation}
\label{m3zero}
\hat{\xi} = 3 (A - A_Q) + \mathcal{O}((A-A_Q)^2).
\end{equation}

In order to use (\ref{matching-condition-2}), we need to compute the coefficient $F(3)$ in (\ref{constant-D})
from a linear algebraic system. From (\ref{polynomial-2}) with $n = 0$, (\ref{dominant-u-C-n}), and (\ref{dominant-R-2-n}),
we obtain
$$
u_1(r) = 12 \left[ r^{-\frac{1}{2}} - \frac{4}{3} + \frac{1}{3} r^{\frac{1}{2}} \right]
$$
and
$$
R_2(r) = 12^2 \left[ -2 r^{-\frac{3}{2}} + 2 r^{-\frac{1}{2}} - \frac{2}{3} \right].
$$
From (\ref{linear-ode-x}) with $m = 3$ and $n = 0$ we have:
$$
L r^{\frac{1}{2}} = 4, \quad L r^0 = 1 + \frac{3}{2} r^{-\frac{1}{2}}, \quad L r^{-\frac{1}{2}} = 2 r^{-\frac{1}{2}}, \quad L r^{-\frac{3}{2}} = 4 r^{-\frac{3}{2}},
$$
and we can find a unique solution of the linear equation (\ref{inhomogeneous-eq}) in the form
$$
u_2(r) = 12^2 \left[ -\frac{1}{2} r^{-\frac{3}{2}} + r^{-\frac{1}{2}} - \frac{1}{6} r^{\frac{1}{2}} \right],
$$
from which $F(3) = -72$. The matching condition (\ref{matching-condition-2}) yields
$$
-\frac{1}{4} \hat{\xi}^2 = \frac{1}{4} \alpha (3-m) - 72 \alpha^2 + \mathcal{O}(\alpha (3-m)^2, \alpha^2 (3-m), \alpha^3).
$$
Substituting $\hat{\xi} = - 12 \sqrt{2} \alpha + \mathcal{O}(\alpha^2)$, we obtain
\begin{equation}
\label{m3inf}
3-m = \mathcal{O}(\hat{\xi}^2).
\end{equation}
Although the approximation (\ref{m3inf}) is not definite due to the cancelation of the linear term in $\hat{\xi}$,
we will show numerically in \S\ref{numerics} that the dependence of $3 - m$ is indeed quadratic with respect to $\hat{\xi}$,
see figure \ref{local_to_3}(a). The precise constant of this quadratic dependence can only be computed if the outer expansion (\ref{outer})
is expanded to next order $\mathcal{O}(\alpha^3)$, which is not computed here. We also see on
figure \ref{local_to_3}(b) that the approximation (\ref{m3zero}) agrees well with the numerical results.

\subsubsection{Behaviour local to $m = 5$:}

On setting $k=3$ and $n = 0$ in (\ref{constant-B}) and (\ref{constant-E-exact}), we obtain $B'(5) = 6$
and $E(5) = -120$. Since $A_Q = 1/\sqrt{3}$, we obtain from (\ref{matching-condition-1}) and (\ref{matching-condition-3}):
$$
\hat{\xi} =120 \sqrt{3} \alpha + \mathcal{O}(\alpha^2), \quad \frac{A_Q}{A} = 1 + 81 \alpha + \mathcal{O}(\alpha^2),
$$
that is,
\begin{equation}
\label{m5zero}
\hat{\xi} = -\frac{40}{9} (A - A_Q) + \mathcal{O}((A-A_Q)^2).
\end{equation}
The coefficient $F(5)$ in (\ref{constant-D}) is computed from a linear algebraic system.
From (\ref{polynomial-3}) with $n = 0$, (\ref{dominant-u-C-n}), and (\ref{dominant-R-2-n}), we obtain
$$
u_1(r) = -120 \left[ r^{-\frac{2}{3}} - \frac{9}{4} r^{-\frac{1}{3}} + \frac{27}{20} - \frac{9}{40} r^{\frac{1}{3}} \right]
$$
and
$$
R_2(r) = 120^2 \left[ -\frac{15}{2} r^{-\frac{5}{3}} + \frac{207}{16} r^{-\frac{4}{3}} - \frac{189}{20} r^{-\frac{2}{3}}
+ \frac{81}{16} r^{-\frac{1}{3}} - \frac{243}{320} \right].
$$
Thanks to the algebra obtained from (\ref{linear-ode-x}) for $m = 5$ and $n = 0$:
\begin{eqnarray*}
& \phantom{t} &
L r^{\frac{1}{3}} = 6, \quad L r^0 = 1 + \frac{10}{3} r^{-\frac{1}{3}}, \quad
L r^{-\frac{1}{3}} = 2 r^{-\frac{1}{3}} + \frac{4}{3} r^{-\frac{2}{3}}, \\
& \phantom{t} &
L r^{-\frac{2}{3}} = 3 r^{-\frac{2}{3}}, \quad
L r^{-\frac{4}{3}} = 5 r^{-\frac{4}{3}} - \frac{2}{3} r^{-\frac{5}{3}}, \quad
L r^{-\frac{5}{3}} = 6 r^{-\frac{5}{3}},
\end{eqnarray*}
we can find a unique solution of the linear equation (\ref{inhomogeneous-eq}) in the form
$$
u_2(r) = 120^2 \left[ -\frac{77}{80} r^{-\frac{5}{3}} + \frac{207}{80} r^{-\frac{4}{3}}
-\frac{387}{80} r^{-\frac{2}{3}} + \frac{243}{64} r^{-\frac{1}{3}} - \frac{243}{320} \right],
$$
from which $F(5) = -13860$. The matching condition (\ref{matching-condition-2}) yields
$$
-\frac{1}{3} \hat{\xi}^2 = \frac{2}{9} \alpha (m-5) - 13860 \alpha^2 + \mathcal{O}(\alpha (m-5)^2, \alpha^2 (m-5), \alpha^3),
$$
from which we obtain
\begin{equation}
\label{m5inf}
5-m = \frac{27 \sqrt{3}}{4} \hat{\xi} + \mathcal{O}(\hat{\xi}^2).
\end{equation}
The asymptotic dependencies (\ref{m5zero}) and (\ref{m5inf}) will be compared with the numerical data
in \S \ref{numerics}, where we will see the excellent agreement between them, see figure \ref{local_to_5}(a,b).

\subsubsection{\label{localm7}Behaviour local to $m = 7$}

On setting $k=4$ and $n = 0$ in (\ref{constant-B}) and (\ref{constant-E-exact}), we obtain
$B'(7) = -72$ and $E(7) = 1680$. Since $A_Q = 1/2$, we obtain from (\ref{matching-condition-1}) and (\ref{matching-condition-3}):
$$
\hat{\xi} = -3360 \alpha + \mathcal{O}(\alpha^2), \quad \frac{A_Q}{A} = 1 + 1024 \alpha + \mathcal{O}(\alpha^2),
$$
that is,
\begin{equation}
\label{m7zero}
\hat{\xi} = \frac{105}{16} (A - A_Q) + \mathcal{O}((A-A_Q)^2).
\end{equation}
The coefficient $F(7)$ in (\ref{constant-D}) is computed from a linear algebraic system.
From (\ref{polynomial-4}) with $n = 0$, (\ref{dominant-u-C-n}), and (\ref{dominant-R-2-n}), we obtain
$$
u_1(r) = 1680 \left[ r^{-\frac{3}{4}} - \frac{16}{5} r^{-\frac{2}{4}} + \frac{16}{5} r^{-\frac{1}{4}}
- \frac{128}{105} + \frac{16}{105} r^{\frac{1}{4}} \right]
$$
and
$$
R_2(r) = 1680^2 \left[ -\frac{84}{5} r^{-\frac{7}{4}} + \frac{1456}{25} r^{-\frac{6}{4}}
- \frac{1504}{25} r^{-\frac{5}{4}}
+ \frac{192}{5} r^{-\frac{3}{4}} - \frac{13568}{525} r^{-\frac{2}{4}} + \frac{512}{75} r^{-\frac{1}{4}} - \frac{1024}{1575} \right].
$$
Thanks to the algebra obtained from (\ref{linear-ode-x}) for $m = 7$ and $n = 0$:
\begin{eqnarray*}
& \phantom{t} &
L r^{\frac{1}{4}} = 8, \quad L r^0 = 1 + \frac{21}{4} r^{-\frac{1}{4}}, \quad
L r^{-\frac{1}{4}} = 2 r^{-\frac{1}{4}} + 3 r^{-\frac{2}{4}}, \quad
L r^{-\frac{2}{4}} = 3 r^{-\frac{2}{4}} + \frac{5}{4} r^{-\frac{3}{4}}, \\
& \phantom{t} &
L r^{-\frac{3}{4}} = 4 r^{-\frac{3}{4}}, \quad
L r^{-\frac{5}{4}} = 6 r^{-\frac{5}{4}} - r^{-\frac{6}{4}}, \quad
L r^{-\frac{6}{4}} = 7 r^{-\frac{6}{4}} - \frac{3}{4} r^{-\frac{7}{4}}, \quad
L r^{-\frac{7}{4}} = 8 r^{-\frac{7}{4}},
\end{eqnarray*}
we can find a unique solution
$$
u_2(r) = 1680^2 \left[ -\frac{509}{350} r^{-\frac{7}{4}} + \frac{3616}{525} r^{-\frac{6}{4}}
- \frac{752}{75} r^{-\frac{5}{4}} + \frac{4376}{315} r^{-\frac{3}{4}} - \frac{2898}{211} r^{-\frac{2}{4}}
+ \frac{128}{25} r^{-\frac{1}{4}} -\frac{1024}{1575} \right],
$$
from which $F(7) = -4104576$. The matching condition (\ref{matching-condition-2}) yields
$$
-\frac{3}{8} \hat{\xi}^2 = \frac{9}{32} \alpha (7-m) - 4104576 \alpha^2 + \mathcal{O}(\alpha (7-m)^2, \alpha^2 (7-m), \alpha^3),
$$
from which we obtain
\begin{equation}
\label{m7inf}
7-m = \frac{2048}{15} \hat{\xi} + \mathcal{O}(\hat{\xi}^2).
\end{equation}
The asymptotic dependencies (\ref{m7zero}) and (\ref{m7inf}) will be compared with the numerical data
in \S\ref{numerics}, where we will see the excellent agreement between them, see figure \ref{local_to_7}(a,b).

\section{\label{numerics}Numerical results}

Here we employ numerical methods to verify the analytical results obtained in \S\ref{matching_sec}.
A fit scheme must be capable of furnishing numerical solutions to the ODE (\ref{ode}) which connect the far-field and near-field
behaviours given by (\ref{ocean}) and either (\ref{onebeh}) for $\hat{\xi}_- > 0$ or (\ref{genes}) for $\hat{\xi}_- < 0$.
Such a numerical method has already been presented in \cite{FosterPel} and so in the interests of brevity
we will give a short summary --- interested readers are referred to \cite{FosterPel} for full details.

Finding numerical approximations of solutions for $H_-$ is a problem that can be tackled using a shooting technique. An appropriate shooting parameter is the value of $A$ in the far-field behaviour (\ref{ocean}). On selecting a value for $A$ the behaviour (\ref{ocean}) can be used to define approximate initial data for $H_-(\xi)$ at some very large, yet finite, value of $\xi$ to begin numerical integration of the ODE (\ref{ode}) in the direction of decreasing $\xi$. The integration can be continued until either the value of $H_-(\xi)$
or $H_-^m(\xi) H'_-(\xi)$ vanishes at some $\xi = \hat{\xi}$; it was proven in \cite{FosterPel} that at least one of these two conditions will be reached
for the unique solution satisfying (\ref{ocean}). The solution to (\ref{ode}) we are looking for
satisfy both of the aforementioned conditions. It is by iterating on the value of $A$ that proper solution(s) for $H_-$ can be found that satisfy both
$H_-(\hat{\xi}_-)=0$ and $H^m_-(\hat{\xi}_-) H'_-(\hat{\xi}_-)=0$, for some $\hat{\xi}_-$.
We denote the special value(s) of $A$ that give rise to solutions satisfy these conditions by $A^*$.

Values of $A^*$ as a function of $m$ are shown in figure \ref{all-sols-negative-t} when the exponent $n$ of the absorption term
in the slow diffusion equation (\ref{heat}) is set equal to $1/2,0,-1/2,-1$. From these figures we observe that
at each value of $m=m_k=(2k-1)(1-n)$ for $k \in \mathbb{N}$ two branches of solutions (one with $\hat{\xi}_->0$ (red)
and the other with $\hat{\xi}_- < 0$ (blue)) bifurcate from the main black branch corresponding to
the exact solution (\ref{exact-solution}) with $\hat{\xi}_- =0$.
The bifurcation points were identified in \S\ref{proplinop}, see (\ref{bifurcation-value}).
Moreover, the alternation of the red and blue curves above the black curve between adjacent bifurcations
observed on figure \ref{all-sols-negative-t} can be explained by the sign alternation of $E(m_k)$ in
(\ref{constant-E-exact}) and (\ref{matching-condition-1}) between two values of $k$.

\begin{figure} \centering
\includegraphics[width=0.45\textwidth]{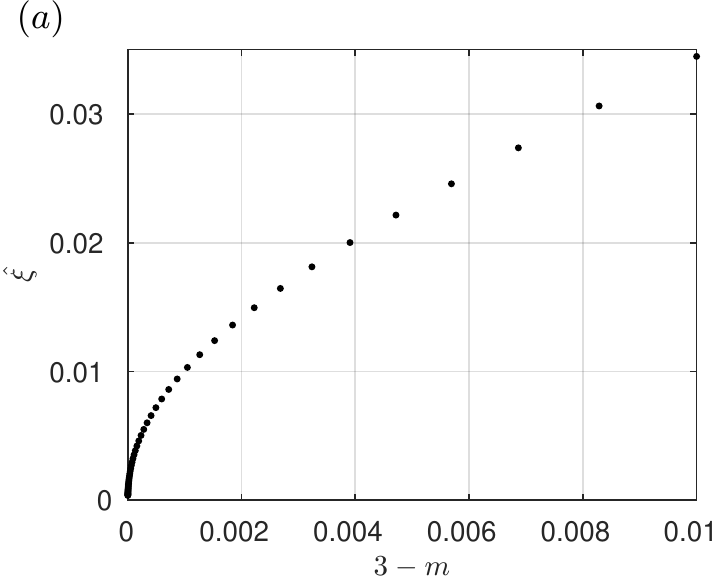} \hspace{2em}
\includegraphics[width=0.45\textwidth]{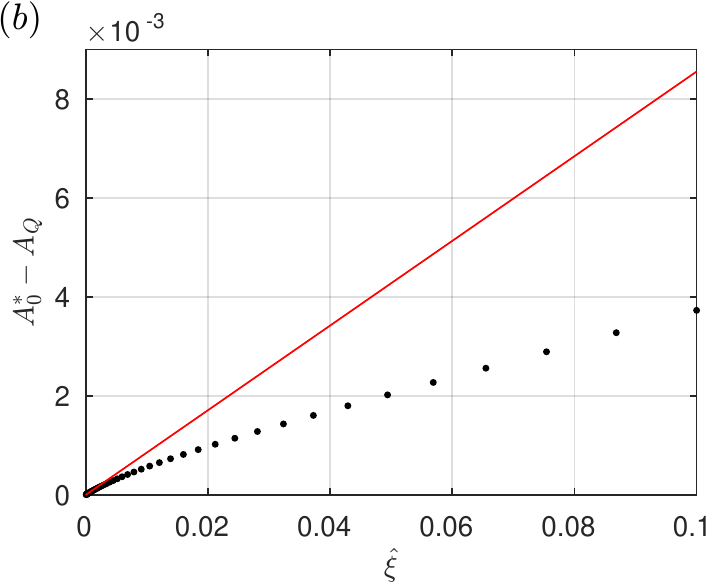}
\caption{Panel (a): The variation of $\hat{\xi}$ with $3-m$ as predicted by shooting. The observed quadratic dependence of $3 - m$ on $\hat{\xi}$ is in agreement with (\ref{m3inf}). Panel (b): The variation of $A-A_Q$ with $\hat{\xi}$ local to $m=3$ as predicted by both shooting (black dots) and (\ref{m3zero}) (red line).}
\label{local_to_3}
\end{figure}

\begin{figure} \centering
\includegraphics[width=0.46\textwidth]{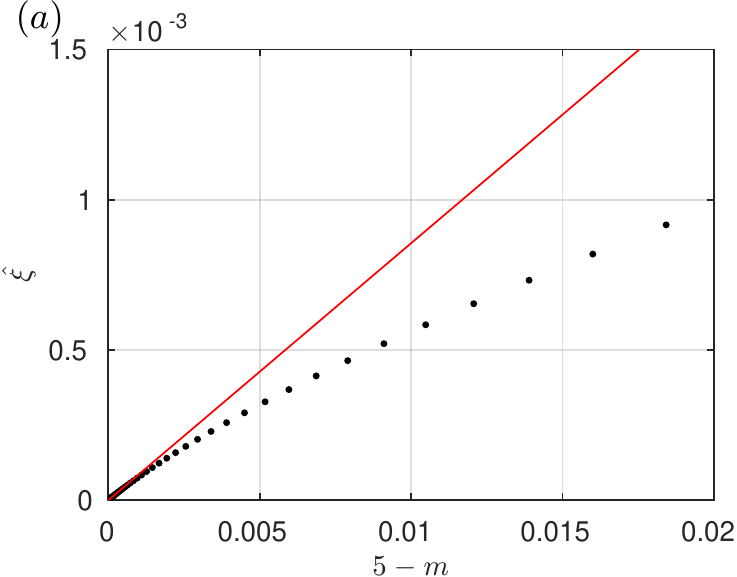} \hspace{2em}
\includegraphics[width=0.44\textwidth]{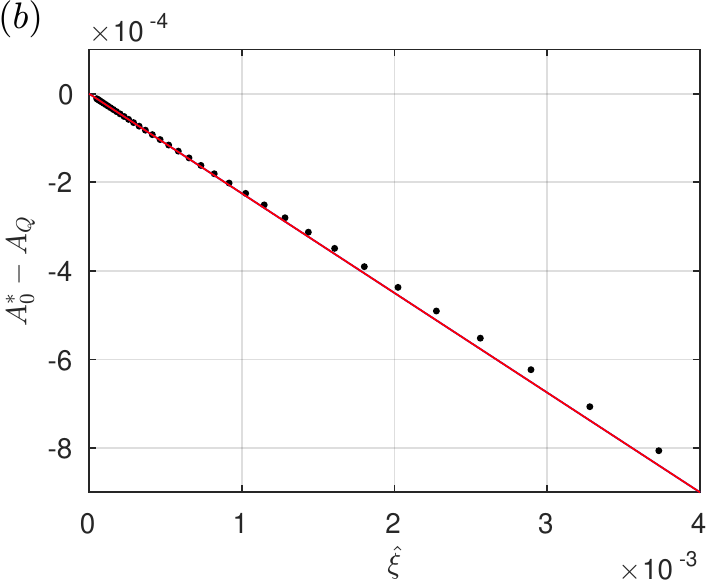}
\caption{Panel (a) shows the variation of $\hat{\xi}$ with $5-m$ and panel (b) shows the variation of $A-A_Q$ with $\hat{\xi}$ local to $m=5$.
Black dots indicate numerical results
whereas the red lines in panels (a) and (b) are the behaviours predicted by (\ref{m5zero}) and (\ref{m5inf}) respectively.}
\label{local_to_5}
\end{figure}

\begin{figure} \centering
\includegraphics[width=0.45\textwidth]{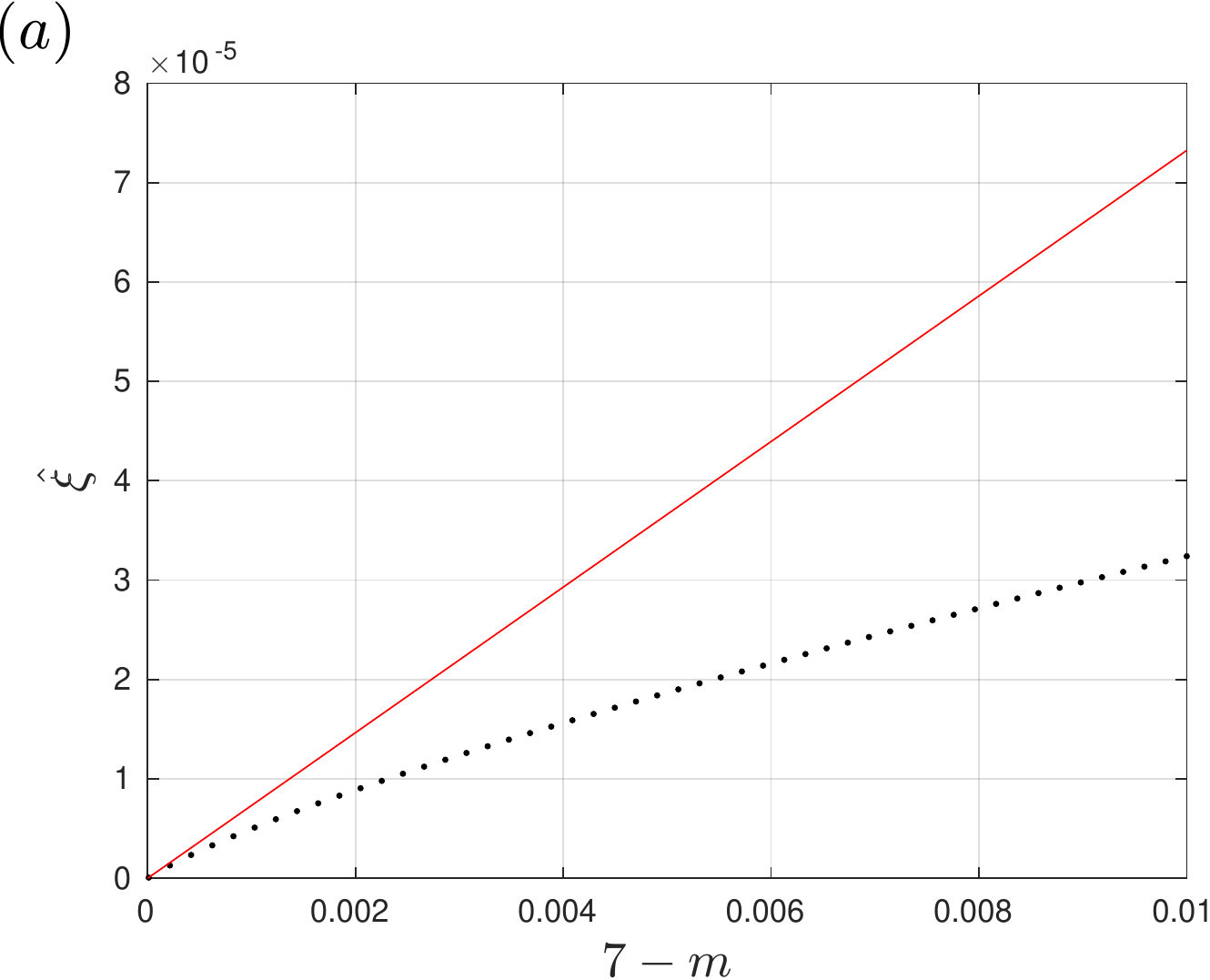} \hspace{2em}
\includegraphics[width=0.45\textwidth]{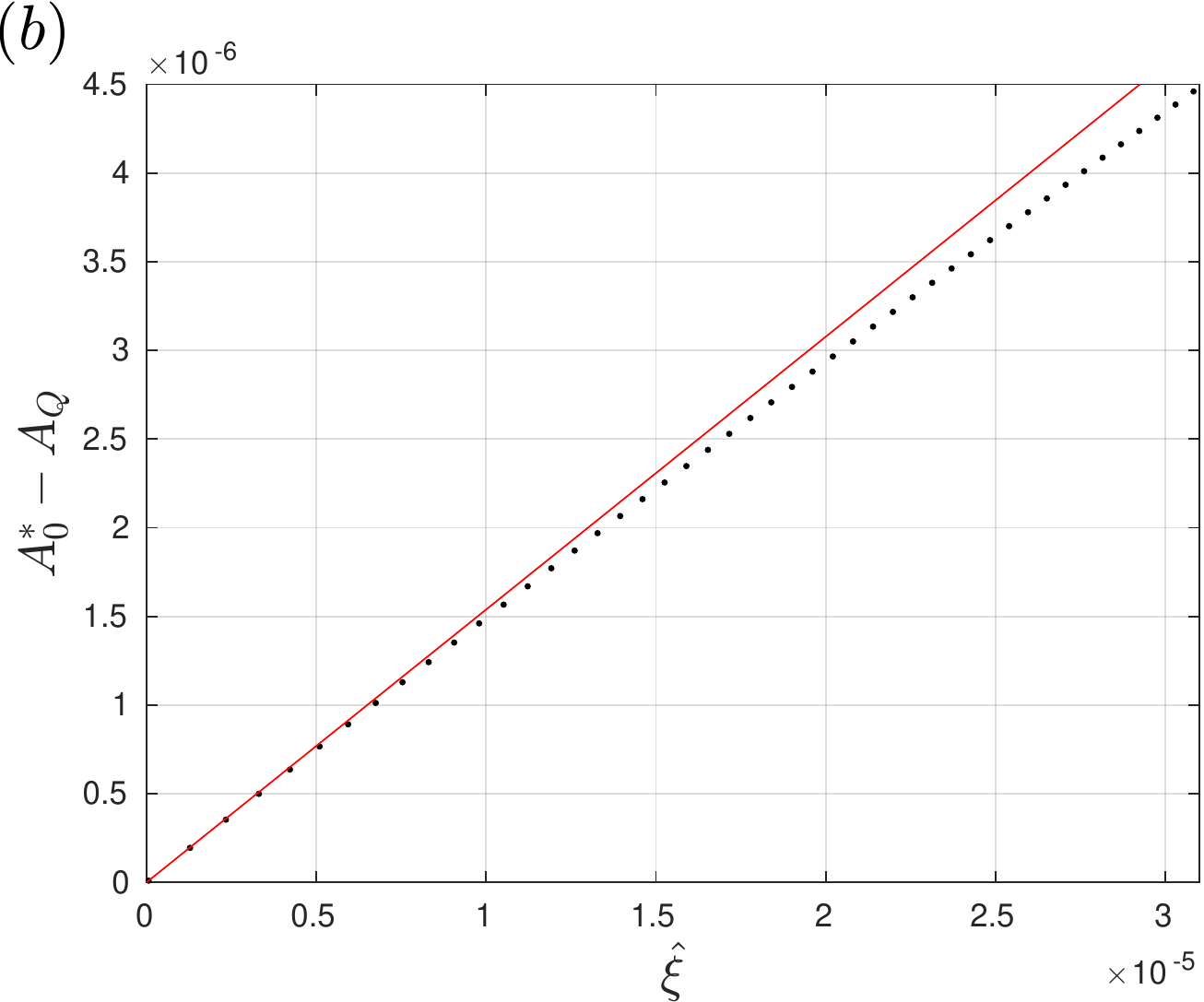}
\caption{Panel (a) shows the variation of $\hat{\xi}$ with $7-m$ and panel (b) shows the variation of $A-A_Q$ with $\hat{\xi}$ local to $m=7$.
Black dots are numerical results
whereas the red lines in panels (a) and (b) are the behaviours predicted by (\ref{m7zero}) and (\ref{m7inf}) respectively.}
\label{local_to_7}
\end{figure}

The matched asymptotics analysis in \S\ref{matched-asymptotics} yields prediction of the local
dependencies of both $\hat{\xi}_-$ and $A^*-A_Q$ on $m_k - m$ local to each bifurcation point
$m=m_k$ for $n = 0$. In figures \ref{local_to_3}-\ref{local_to_7} we show a comparison between
these predictions and the results of the numerical shooting scheme outlined above. We observe
excellent agreement in all cases, thereby supporting both the analysis presented here and
the accuracy of the numerical scheme proposed in \cite{FosterPel}.

\section{Conclusions}
\label{conclusion}

We conclude this paper by placing the self-similar solutions into the original physical context,
\emph{i.e.}, in terms of the PDE (\ref{heat}). On transforming to travelling wave-type coordinate system that moves with the position of the left-hand interface (with position $x=s(t)$), using the change of variables $\eta = x - s(t)$, and seeking asymptotic solutions to the PDE (\ref{heat})
for small values of the moving coordinate, $\eta$, we find solutions with local behavior in (\ref{first-wave})
for $\dot{s} < 0$ and in (\ref{second-wave}) for $\dot{s} > 0$.
The local behaviour in (\ref{first-wave}) is termed an advancing interface, since its motion acts
to enlarge the domain of compact support, whereas the local behaviour in (\ref{second-wave})
is termed a receding solution. Examining (\ref{first-wave}) we see that the advancing wave is largely controlled by the exponent $m$ and the diffusive-type term in (\ref{heat}). Physically this corresponds to the forward motion of an interface being driven by fluid pressure ($m=3$), a biological population pressure ($m=2$) or nonlinear heat conduction ($m=4$). Contrastingly, the alternative behaviour, (\ref{second-wave}), is controlled by the absorption term in (\ref{heat}) and relates the physical mechanisms of fluid evaporation or absorption into a substrate ($n=0$ or $n=1$), the constant death rate of a biological population ($n=0$) or the loss of heat.

The physical motivation behind the current work is elucidating the processes by which the former local solution becomes the latter (a reversing solution) or vice versa (an anti-reversing solution). In addition to (anti-)reversing solutions we have also studied solutions which advance (recede) instantaneously halt/pause at $t=0$ and then continue advancing (receding) corresponding to the former (latter) behaviour for both negative and positive $t$. Reversing solutions are manifested in the self-similar context by solutions for $H_\pm$ with $\hat{\xi}_\pm>0$. On referring to figure \ref{all-sols-negative-t} we see that at least one such solution exist for all $m \gtrsim m_2$. Contrastingly, it seems that anti-reversing solutions, corresponding to $\hat{\xi}_\pm<0$, exist for $m_1 < m < m_2$. The receding pausing ($\hat{\xi}_->0$ and $\hat{\xi}_+<0$) solutions emerge from the primary (black) branch at $m=m_3$ and persist for $m>m_3$. A branch of advancing pausing solutions ($\hat{\xi}_->0$ and $\hat{\xi}_+<0$)  also emerge from the main branch at this point but persist only for a much smaller range of $m$.

\vspace{0.5cm}

\noindent{\bf Acknowledgements.}
J.F. was supported by a postdoctoral fellowship at McMaster University.
He thanks B. Protas for hospitality and many useful discussions.
The work of P.G. was performed within an undergraduate research project in physics
at McMaster University.
The authors are also grateful to A. D. Fitt for inspiration and useful discussions.

\end{document}